\documentclass[fleqn,usenatbib]{mnras}

\usepackage{newtxtext,newtxmath}
\usepackage{graphicx}% Include figure files
\usepackage{dcolumn}% Align table columns on decimal point

\usepackage[usenames,dvipsnames]{xcolor}
\hypersetup{colorlinks=true,citecolor=blue,linkcolor=OliveGreen}

\usepackage{soul}%highlighting {blah}

\usepackage{subfigure}
\usepackage{wasysym}
\usepackage{verbatim}
\usepackage{color}
\usepackage{mathtools}
\usepackage{hyperref}
\usepackage{soul}
\usepackage{pbox}
\usepackage[export]{adjustbox}

\usepackage{slashed}
\usepackage{multirow}
\usepackage{booktabs}  % for prettier-than-default tables
\usepackage{natbib}
\usepackage{float}
\usepackage{afterpage}
\usepackage{dcolumn}
\usepackage{bm}
\usepackage{amssymb,amsmath}  
\usepackage{lipsum}
\usepackage{setspace}
\usepackage{etoolbox}

\def\beq{\begin{equation}}
\def\eeq{\end{equation}}

\def\mnras{Mon. Not. R. Astron. Soc}
\def\physrep{Physics Reports}

\usepackage[T1]{fontenc}
\usepackage{ae,aecompl}

\usepackage{graphicx}	% Including figure files
\usepackage{amsmath}	% Advanced maths commands
\usepackage{amssymb}	% Extra maths symbols

\title[Extensions to models of the galaxy-halo connection]{Extensions to models of the galaxy-halo connection}

\author[B. Hadzhiyska et al.]{
Boryana Hadzhiyska,$^{1}$\thanks{E-mail: boryana.hadzhiyska@cfa.harvard.edu}
Sownak Bose,$^{1}$
Daniel Eisenstein,$^{1}$
\ Lars Hernquist$^{1}$
\\
$^{1}$Harvard-Smithsonian Center for Astrophysics, 60 Garden St., Cambridge, MA 02138, USA\\
}
\date{Accepted XXX. Received YYY; in original form ZZZ}

\pubyear{2019}

\begin{document}
\label{firstpage}
\pagerange{\pageref{firstpage}--\pageref{lastpage}}
\maketitle

\begin{abstract}
We explore two widely used empirical models for the galaxy-halo connection, subhalo abundance matching (SHAM) and the halo occupation distribution (HOD) and compare their predictions with the state-of-the-art hydrodynamical simulation IllustrisTNG (TNG) for a range of statistics that quantify the galaxy distribution at $n_{\rm gal}\approx1.3\times10^{-3}\,[{\rm Mpc}/h]^{-3}$. We observe that in their most straightforward implementations, both models fail to reproduce the two-point clustering measured in TNG.  We find that SHAM models constructed using the relaxation velocity, $V_{\rm relax}$, and the peak velocity, $V_{\rm peak}$, perform best, and match the clustering reasonably well, although neither model captures adequately the one-halo clustering. Splitting the total sample into sub-populations, we discover that SHAM overpredicts the clustering of high-mass, blue, star-forming, and late-forming galaxies and uderpredicts that of low-mass, red, quiescent, and early-forming galaxies. We also study various baryonic effects, finding that subhalos in the dark-mater-only simulation have consistently higher values of their SHAM-proxy properties than their full-physics counterparts. We then consider a two-dimensional implementation of the HOD model augmented with a secondary parameter (environment, velocity anisotropy, $\sigma^2R_{\rm halfmass}$, and total potential) and tuned so as to match the two-point clustering of the IllustrisTNG galaxies on large scales. We analyze these galaxy populations adopting alternative statistical tools such as galaxy-galaxy lensing, void-galaxy cross-correlations and cumulants of the smoothed density field, finding that the hydrodynamical galaxy distribution disfavors $\sigma^2 R_{\rm halfmass}$ and the total potential as secondary parameters, while the environment and velocity anisotropy samples are consistent with full-physics across all statistical probes examined. Our results demonstrate the power of examining multiple statistics that characterize the galaxy field, which enables us to determine a hierarchy in the efficacy of secondary parameters for augmenting the galaxy-halo connection.
\end{abstract}


%%%%%%%%%%%%%%%%%%%%%%%%%%%%%%%%%%%%%%%%%%%%%%%%%%

%%%%%%%%%%%%%%%%% BODY OF PAPER %%%%%%%%%%%%%%%%%%

\section{Introduction}

The study of galaxy clustering provides a 
wonderful opportunity to learn about
galaxy formation and evolution as
well as the fundamental laws governing our 
cosmological model. Small-scale clustering
offers information regarding the relationship between galaxies
and their dark matter (DM) halo hosts and also stellar
population properties, whereas large-scale clustering
of galaxies yields insight into many of the unresolved  
mysteries of our Universe such as the nature of dark
matter and dark energy and its cosmological makeup and
early history \citep{1980lssu.book.....P}. 
Computing correlation function statistics
of different physical quantities at different
scales allows for the 
breaking of multiple degeneracies among astrophysical
parameters and thus can yield tighter constraints on those and, eventually, the cosmological parameters themselves. 

Extracting information from galaxy clustering, however,
relies on the accuracy of predictions 
from galaxy formation models and our cosmological theories. 
On small scales, this task is challenging because one
enters the non-linear regime, where analytical solutions
are not available, and stochastic processes
such as mergers, tidal stripping, star formation, and
non-linear gravitational collapse take place. These processes are best mimicked
and studied in numerical simulations of which there are two
types \citep[see][for a review]{{2012PDU.....1...50K}}.
The first kind implements jointly the evolution of the
DM and baryon fluid together with recipes for unresolved
astrophysical processes such as feedback from supernovae
and star formation. These so-called hydrodynamical simulations
yield a direct prediction for
the distribution and properties of galaxies, but the
computational expense for simulating sufficiently large
volumes (i.e. at volumes comparable to future 
galaxy redshift surveys) at an adequate resolution is great 
\citep{2014MNRAS.444.1518V,2015MNRAS.446..521S,2020arXiv200601039L,2020NatRP...2...42V}.
The current box size of these simulations is on the order
of only a few hundred Mpc/$h$.

On the other hand, simulating gravitation-only interactions
is much more inexpensive from a numerical resource point of view.
To correctly mimic the statistics and systematics of 
observational surveys, it is necessary to model galaxy clustering
in simulations with volumes on the order of 1 Gpc$^3/h^3$
\citep{2008MNRAS.387..921A}, 
which are eminently feasible in these DM-only or N-body simulations. 
The downside is that one needs to resort
to ``painting'' galaxies {\it a posteriori}, after the simulation
has been completed. This approach is supported by leading
theories of galaxy formation in which the locations of
galaxies are predominantly determined by the properties
of the DM halos they reside in \citep[see][for a review]{2020NatRP...2...42V}. There are many recipes
one can adopt for populating DM halos with galaxies, but 
unfortunately there is a lot of uncertainty surrounding
these choices, as the relationship between halos
and galaxies is not straightforward \citep[see][for a review]{{2018ARA&A..56..435W}}.

The galaxy-halo relation is at the crux of many of the standard techniques
for ``painting'' galaxies on top of halos, including one of the most
widely used and computationally inexpensive 
empirical
(or phenomenological) models known as the halo
occupation distribution (HOD) model. In its simplest manifestation, the HOD framework describes
the number of galaxies residing in a host halo as a function of
halo mass
\citep{2000MNRAS.318.1144P,2000MNRAS.318..203S,2001ApJ...546...20S,HOD}, 
remaining agnostic about any other halo property.
It rests on the long-standing and widely
accepted theoretical prediction that 
halo mass is the attribute that most strongly
influences the halo abundance and halo
clustering as well as the properties of the 
galaxies in it \citep{1978MNRAS.183..341W,1984Natur.311..517B}.
The HOD model is currently a popular choice for building
mock galaxy catalogs and interpreting 
observations of the correlation
function of galaxies, leading to constraints on the typical
halo masses of observed galaxies and on the values of 
cosmological parameters \citep{2019ApJ...874...95Z}.

Another empirical method for populating collisionless
N-body simulations is called the subhalo abundance matching
(SHAM) method \citep[e.g.][]{2007astro.ph..1096V,2006ApJ...647..201C}.
In its original implementation, SHAM assumes
an injective and monotonic relation between galaxies and self-bound DM
structures (subhalos) based on a pair of specified parameters.
Typically one associates the most massive galaxies, in terms
of their stellar mass, with the most massive DM subhalos, where
one often uses velocity-related subhalo properties as mass proxies
such as the maximum circular velocity. To introduce more realism, recent
implementations have also introduced stochasticity into that relation, 
\citep[e.g.][]{2010ApJ...717..379B,2014ApJ...783..118R}.
The thus obtained
galaxies are placed at the centers of the subhalos 
and given the same velocity as their corresponding subhalos. 
SHAM in its original form, thus makes predictions for the
clustering of galaxies, but not for any physical galaxy properties such
as stellar mass, star formation rate, metallicity, etc.
SHAM-derived galaxy catalogs have been shown to reproduce
the observed galaxy clustering over a broad redshift range
across different datasets,
\citep[e.g.][]{2013MNRAS.432..743N,2014ApJ...783..118R,2013MNRAS.436.1142S}. 
On the other hand, comparisons with hydrodynamical
simulations have yielded less conclusive results, depending on the
choice of SHAM subhalo proxy, redshift of interest and choice
of galaxy density \citep[e.g.][]{2012MNRAS.423.3458S,2016MNRAS.460.3100C,2020arXiv200503672C}.
It is important to consider that
any such comparison is strongly dependent on the galaxy formation
and evolution models adopted in the
hydro simulations. The `precision cosmology' goals of future galaxy surveys will place even
more stringent demands regarding the accuracy of the population models employed.

One of the main limitations to improving the accuracy of 
the galaxy clustering measurements and systematics thereof
is the way in which these models
handle ``galaxy assembly bias'', which
is a manifestation of the large-scale discrepancy between the DM
distribution and that of galaxies. Assembly bias refers to details of
the galaxy-halo connection related
to halo assembly history beyond just
present-day halo mass \citep{2007MNRAS.374.1303C}
such as halo formation time, environment, 
concentration, triaxiality, spin, and 
velocity dispersion.
Galaxy assembly bias results from two effects: halo
assembly bias and occupation variation. The former 
manifests itself as a difference in the halo clustering
among halos of the same mass that differ in some secondary
property \citep[e.g. formation time, concentration, spin,][]{2005MNRAS.363L..66G}, while the latter comes from the
dependence of the galaxy population on properties of the halo
host other than its mass \citep{2018ApJ...853...84Z,2018MNRAS.480.3978A}.

The standard (mass-only) implementation of the HOD does not consider halo
properties apart from its mass and hence does not incorporate
galaxy assembly bias effects. On the other hand, the subhalo abundance
matching techniques have an intrinsic dependence on the halo assembly
history -- recently-formed halos tend to have richer substructure, i.e.
more subhalos. However, the baseline
SHAM model disregards baryonic effects on processes internal to halos, such as tidal stripping and subhalo
disruption, which may affect our ability to link subhalos in N-body 
simulations to those in hydro simulations.
The most straightforward versions of both models
also fail to implement a dependence on environmental properties,
which have recently been supported through a growing body of evidence
\citep[e.g.][]{2019arXiv190302007R,2019arXiv190200030M,2020MNRAS.493.5506H}.

There have been several attempts to incorporate assembly bias into
the HOD framework \citep{2015MNRAS.454.3030P,2016MNRAS.460.2552H,2018MNRAS.478.2019Y,2019MNRAS.488.3541W,2019MNRAS.485.1196Z,2019ApJ...872..115V},
most of which have used halo concentration
(or closely related quantities as proxies) as the main halo property.
This parameter has been shown, however, to
be insufficient in reproducing the full galaxy assembly bias signal
\citep{2007MNRAS.374.1303C,2020MNRAS.493.5506H,2020MNRAS.492.2739X}.
Other parameters such as environment have not
been explored as thoroughly in the literature
\citep{2016arXiv160102693M,2020MNRAS.492.2739X}. There has also been
a small number of works attempting to model assembly bias in SHAM, 
e.g. \citet{2017ApJ...834...37L,2020arXiv200503672C}.

Understanding the effect of assembly bias is extremely
important since, if modeled incorrectly, it can lead
to substantial biases in the inferred constraints on cosmological
and galaxy formation properties.
One way to check whether the modeling for future
surveys is done at the required levels of precision
is by testing extensions to these empirical models 
against hydrodynamical simulations.
In this paper, we explore the agreement (or not, as the case may be) between the
state-of-the-art hydrodynamical simulation
IllustrisTNG (TNG) and the
two most widely used empirical models for populating galaxies,
SHAM and HOD. We first study which SHAM parameter choices
exhibit the least amount of discrepancy compared with TNG, delving
into details of the galaxy sub-populations captured best by the
model (e.g. red vs. blue galaxies, satellites vs. centrals, etc.)
and commenting on differences between the subhalo populations in
the DM-only and full-physics simulation runs. Next we consider
a two-dimensional implementation of the HOD model (2D-HOD), first
introduced in \citet{2020MNRAS.493.5506H}, tuned so as to 
reconcile
the differences between the large-scale correlation
function of the model and the hydro simulation.
Finally, we compare other statistical properties of the two
samples such as galaxy-galaxy lensing, void functions, and
cumulants of the smoothed density field, commenting on the
most appropriate secondary parameter choices.

\section{Methods}
\label{sec:meth}
Here we describe the hydrodynamical simulation
employed in this study, IllustrisTNG, and two standard
empirical galaxy population models,
HOD and SHAM, which are commonly used to populate N-body
simulations and are relatively straightforward to
implement and computationally inexpensive to run.
We test the assumptions of these models on the DM-only
IllustrisTNG simulation and compare them with the corresponding
full-physics galaxy sample.
Combining the two datasets allows us to draw 
direct statistical comparisons between the ``truth'' (defined by
the full-physics run) and the two population models.

\subsection{IllustrisTNG}
In this paper, we consider the hydrodynamical simulation
IllustrisTNG (TNG)
\citep{2018MNRAS.475..648P,
2018MNRAS.480.5113M,2018MNRAS.477.1206N,2018MNRAS.475..676S,2019ComAC...6....2N,TNGbim,2019MNRAS.490.3196P,2019MNRAS.490.3234N}. TNG is a suite of
magneto-hydrodynamic cosmological simulations, 
which were carried out using the \textsc{AREPO}
code \citep{2010MNRAS.401..791S} with 
cosmological parameters consistent with the
\textit{Planck 2015} analysis
\citep{2016A&A...594A..13P}.
These simulations feature a series of improvements
compared with their predecessor, Illustris, such as
improved kinetic AGN feedback and galactic wind
models, as well as the inclusion of magnetic fields.

In particular, we will use Illustris-TNG300-1
(TNG300 thereafter), the largest high-resolution
hydrodynamical simulation from the suite. 
The size of its periodic box
is 205 Mpc$/h$ with 2500$^3$ DM particles
and 2500$^3$ gas cells, implying a DM particle
mass of $3.98 \times 10^7 \ {M_\odot}/h$ and
baryonic mass of $7.44 \times 10^6 \ {M_\odot}/h$. 
We also employ its DM-only counterpart, 
TNG300-Dark, which was evolved with the 
same initial conditions and the same number
of dark matter particles ($2500^3$), each with 
particle mass of $4.73 \times 10^7 \ {M_\odot}/h$.
This gives us an 
opportunity to make a halo-by-halo assignment of galaxies
by cross-matching the full-physics and DM-only simulations.
The halos (groups) in TNG are found with a standard 
friends-of-friends (FoF) algorithm with linking length $b=0.2$
run on the dark matter particles, while the subhalos are identified 
using the SUBFIND algorithm \citep{Springel:2000qu}, which detects 
substructure within the 
groups and defines locally overdense, self-bound particle groups.
We analyze the simulations at the final redshift, $z = 0$.

\subsection{SHAM}
We first consider the empirical galaxy population model called
``subhalo abundance matching'' or SHAM.
In its simplest form, it assumes a perfect match
between stellar mass (or luminosity) as obtained
from a hydrodynamical simulation or galaxy survey and a subhalo property output by
the corresponding N-body run. The latter property can simply be
the total DM mass of the subhalo, 
but recently SHAM models have preferentially been 
using velocity-related parameters, as they are
deemed more resilient to tidal stripping
and other disruptive processes
resulting after the initial subhalo infall \citep[e.g.][]{2016MNRAS.459.3040G}.
A growing body of evidence has also suggested that
populating subhalos based on their early-history properties 
(e.g. velocity at time of infall or peak circular velocity)
also leads to a better agreement with observed 
galaxy clustering \citep[e.g.][]{2016MNRAS.460.3100C}. 
Finally, other SHAM 
implementations also include scatter in the mapping
between hydrodynamical and DM-only objects to 
mimic what is empirically seen in observations
\citep{2020arXiv200503672C}.

\subsection{HOD}
The main assumption of the standard HOD model
is that the number of galaxies belonging to a halo is determined
probabilistically by the mass of that halo.
Typically, the HOD model has a functional form with several 
free parameters which describe
the central and satellite mean occupation functions
as a function of mass, $N_{{\rm cen}}(M_{\rm h})$
and $N_{{\rm sat}}(M_{\rm h})$,
respectively. For each halo, the number of
galaxies is then sampled from these distributions.
A widely used example of this functional form was 
introduced in \citet{Zheng:2004id}
\begin{equation}\label{eq:ncen}
    \left< N_{{\rm cen}} (M_{\rm h}) \right> = \frac{1}{2} \left[ 1 + {\rm erf} \left( \frac{\log M_{\rm h}-\log M_{{\rm min}}}{\sigma_{{\log M}}} \right) \right] 
\end{equation}
\begin{equation}\label{eq:nsat}
    \left<N_{{\rm sat}} (M_{\rm h})\right> = 
    \left( \frac{M_{\rm h}-M_{{\rm cut}}}{M_1}
    \right)^\alpha ,
\end{equation}
where $M_{\rm min}$ is the characteristic minimum
mass of halos that host central galaxies,
$\sigma_{\log M}$ is the width of this transition, 
$M_{{\rm cut}}$ is the characteristic cut-off
scale for hosting satellites, $M_1$ is a 
normalization factor, and $\alpha$ is the power-law
slope. As halo mass proxy, $M_{\rm h}$, the definition we adopt here is
$M_{\rm 200m}$, which is the total mass within a sphere
with mean density 200 times the mean density of the
Universe. 

\begin{figure}
\centering  
\includegraphics[width=0.5\textwidth]{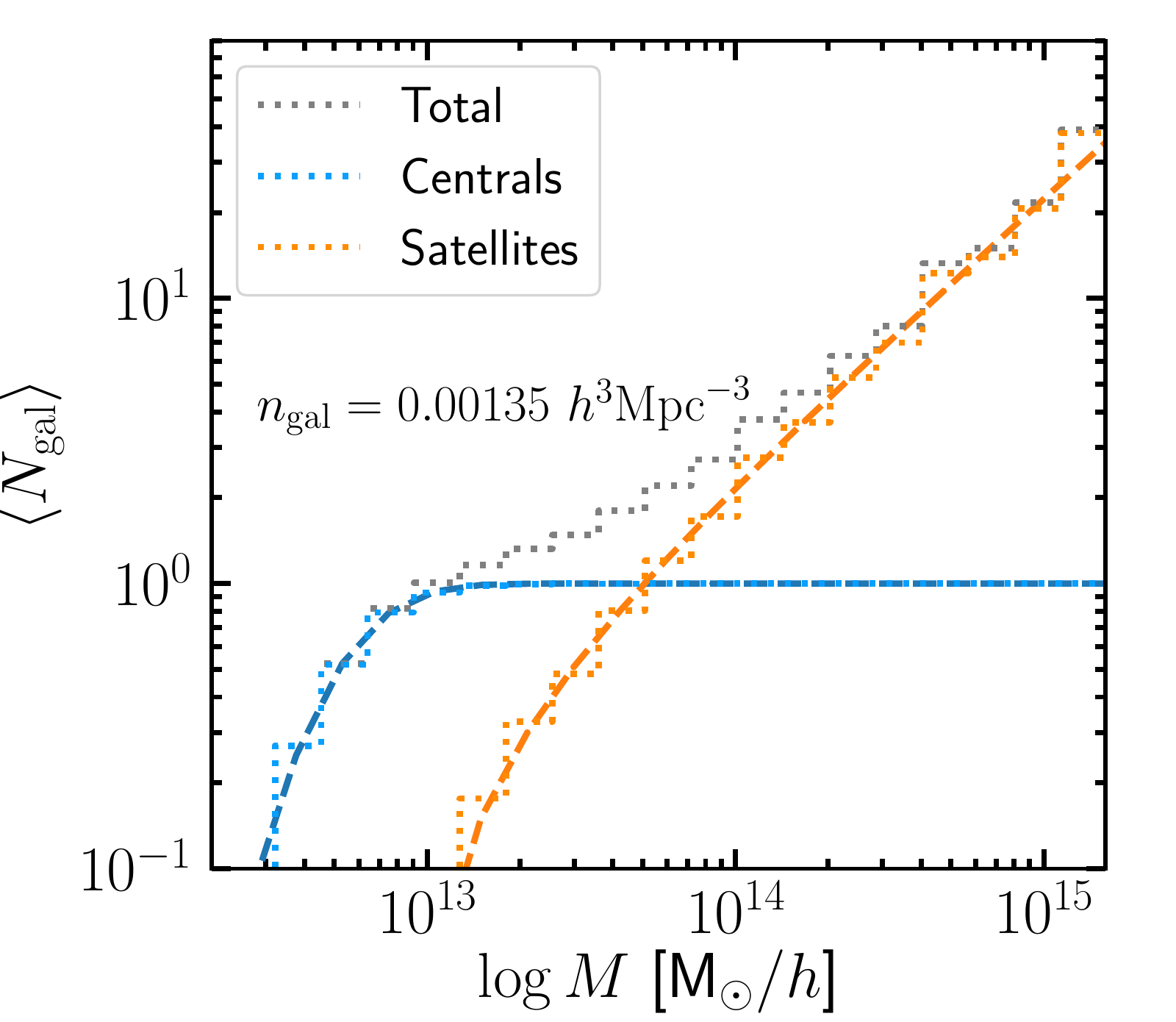}
\caption{Histogram of the mean number of
galaxies per halo
as a function of halo mass (halo occupation distribution)
in TNG300. Here, we use $M_{\rm 200m}$ as the halo mass 
definition and break the galaxy sample
into two populations -- centrals and satellites.
We also show fits (\textit{dashed line}) to these 
populations assuming the 5-parameter HOD 
model described 
in the text \citep{Zheng:2004id}.}
\label{fig:hod_fof}
\end{figure}

In Fig. \ref{fig:hod_fof}, we show the HOD 
derived from TNG300
at $n_{\rm gal} \approx 1.3 \times 10^{-3}
\ [{\rm Mpc}/h]^{-3}$, a galaxy number density
close to the anticipated one in future galaxy surveys.
The figure demonstrates that 
Eqs.~\ref{eq:ncen} and ~\ref{eq:nsat} capture the 
overall shape of the HOD from our simulations very 
well. The corresponding best-fit values for the 5 free 
parameters of this model are:
$\log M_{{\rm min}}=12.712$, 
$\sigma_{{\log M}}=0.287$, $\log 
M_{{\rm cut}}=12.95$, $\log M_1 = 13.62$ and 
$\alpha = 0.98$.

\subsubsection{``Basic'' HOD}
\label{sec:bhod}
Throughout this paper, we will be referring back to the
galaxy sample derived via a mass-only 
prescription of the HOD model,
which we dub ``basic'' HOD. This was introduced
in \citet{2020MNRAS.493.5506H}, where it was found that
the ``basic'' HOD model
leads to a discrepancy of $\sim 15\%$ in the two-point
clustering compared with the hydrodynamical simulation \citep[see also][]{2007MNRAS.374.1303C,2019arXiv190811448B}.
We outline the procedure below:
\begin{itemize}
    \item[1.] Bijectively match as many of the halos across the dark-matter-only (DMO) and full-physics (FP) TNG300-1 simulations (close to
    99\% in the mass range of interest). Keep track
    of how many galaxies the corresponding DM-only
    halos would receive from this match.
    \item[2.] 
    Split the halos into mass bins such that
    the fractional change within each is $\frac{M_{\rm max}-M_{\rm min}}{M_{\rm avg}} \lesssim 5\%$.
    \item[3.] Order the DM-only halos by mass
    and, within each mass bin,
    reassign the number of galaxies by randomly shuffling them. We exclude the 100 most massive halos because the 
    halo mass function contains very few examples of such high-mass systems.
    \item[4.] 
    The galaxies within a given halo are assigned to the
    subhalos in descending order of their $V_{\rm peak}$ velocity 
    (peak magnitude of the circular velocity attained by the subhalo
    at any point in its evolution). The $V_{\rm peak}$ assignment has been shown to provide a good match to the radial distribution of satellites \citep[see][]{sownak}.
\end{itemize}

Fig. \ref{fig:hod_fof} remains
unchanged after performing 
a shuffling of the occupation numbers in 5\% mass
bins following the recipe described above, so
any deviations from the
full-physics galaxy samples are suggestive of 
violations of some of the HOD model assumptions.

\subsubsection{``Fitted'' 2D-HOD}
\label{sec:2dhod}

As shown in \citet{2020MNRAS.493.5506H}, the parameters
that are most influential in shifting the large-scale clustering
in the direction of the hydro simulation result are the local 
environment parameter, $f_{\rm env}$,                  
the velocity anisotropy,                        
$\beta$, and the dynamical mass proxy, $\sigma^2 R_{\rm halfmass}$,
which we will review in Section \ref{sec:pars}.
We do this assuming a perfect association of ranks between 
the halo occupation and the second parameter (e.g. the halo with
the largest value of $f_{\rm env}$ gets assigned the largest number 
of galaxies in each 5\% mass bin). However,
in order to recover the clustering of galaxies in TNG300, 
we need to introduce a scatter into this relation, effectively
obtaining a two-dimensional HOD (2D-HOD) prescription.

Here we introduce a reordering procedure which, like
the ``basic'' HOD model described in Section
\ref{sec:bhod}, preserves the occupation numbers of the
full-physics halos, but matches the clustering of the 
full-physics galaxies on large-scales \citep[see also][]{2020MNRAS.493.5506H}.
This is accomplished by introducing scatter between the 
halo occupation number within each mass bin and a halo parameter
of interest, $p_{\rm match}$.

For each 5\%-mass bin, which contains $N_{\rm h}$ halos,
\begin{itemize}                                                         
  \item[1.] We choose a correlation parameter, $r$, between 0 and 1 and draw 
  $N_{\rm h}$ pairs of $(x,y)$ values from a joint Gaussian distribution
  with mean $[0,0]$ and covariance matrix $[[1,r],[r,1]]$. 
  \item[2.] We convert the array of $\{(x,y)\}$ pairs into an array of pairs
  of integers by assigning an integer value to each entry
  of both $\{x\}$ and $\{y\}$, $\{(x,y)\} \rightarrow \{(i,j)\}$, 
  where $i$ and $j$ are integers between 1 and $N_{\rm h}$
  determined by the rank in descending order of each
  of the elements in the original $\{x\}$ and $\{y\}$ arrays, respectively.
  \item[3.] 
  We form two new arrays: one containing the occupation number of
  the halos in that mass bin, $\{N_{\rm gal}\}$,
  and one containing the values of the
  halo property of interest, $\{p_{\rm match}\}$.
   \item[4.] 
   We then directly associate the drawn array of $\{(i,j)\}$ 
   pairs with the two halo property arrays, 
   so that for a given pair $(\tilde i,\tilde j)$ 
   the halo with $\tilde j^{\rm th}$ highest 
   parameter value $p_{\rm match}$ receives the $\tilde i^{\rm th}$ highest halo occupation number in that bin.
\end{itemize} 

To obtain the amount of  correlation, $r$, between the number
of galaxies per halo and the halo property of choice,
$p_{\rm match}$, we minimize the $\chi^2$-value
of the auto-correlation function, $\xi(r)$, on large
scales between the 2D-HOD derived sample
and the hydrodynamical galaxy sample
\begin{equation}
    \chi^2 = \sum_i \frac{(\xi_{i, \ \rm TNG300} - \xi_{i, \ \rm 2D-HOD})^2}{\sigma_{i, \ \xi}^2},
\end{equation}
where $i$ varies between 1 and the number of 
radial bins and $\sigma_{i, \ \xi}$ is the
jackknife error of the two-point
correlation function of the full-physics sample
in the $i^{\rm th}$ radial bin. We tune the
clustering for 10 radial bins in the range of 1 Mpc$/h$
to 20 Mpc$/h$.

\subsection{Jackknife errors}
\label{sec:jack}
To obtain a measure of the errors associated with 
our various statistical tools (correlation
functions, void size histograms, density field moments),
we split the volume of the simulation into $3^3 = 27$ equal
parts and define 27 subsamples by excluding in each a 
different cube of side $(205/3) 
{\rm \ Mpc}/h \approx 68{\rm \ Mpc}/h$ from the total volume.
We then estimate the statistics of interest for 
the galaxy samples in each of the 27 subsamples and compute
their mean and standard deviation.

As an example, let us consider the estimation of errorbars
in the galaxy clustering. We first
compute the correlation functions in
each of the 27 subsamples using the 
Landy-Szalay estimator \citep{1993ApJ...412...64L}
\begin{equation}
    \xi_{\rm LS}(r) = \frac{DD(r)}{RR(r)}-1 ,
\end{equation} 
assuming periodic boundary conditions.
To obtain the correlation function and corresponding errors
for the full box, we calculate the mean 
and jackknife errors of the correlation 
functions (and their ratios) adopting the standard equations
\begin{equation}
    {\bar \xi(r)}=\frac{1}{n}\sum_{i=1}^{n} {\xi}_i(r)
    \end{equation}
    \begin{equation}
    {\rm Var}[{\xi(r)}]=\frac{n-1}{n} \sum_{i=1}^{n} ({\xi_i(r)} - {\bar \xi(r)})^2 ,
\end{equation}
where $n=27$ and ${\xi_i(r)}$ is the correlation
function value at distance $r$ for subsample $i$ (i.e. excluding the
galaxies residing within volume element $i$ in the correlation
function computation).

\section{SHAM modeling}
The analysis performed in this section
uses the 12000 most massive subhalos (defined by their stellar mass) in
the TNG300 simulation (resulting in a number density of 
$n_{\rm gal} \approx 1.3 \times 10^{-3}
\ [{\rm Mpc}/h]^{-3}$). In this way, we attempt to
limit our study to well-resolved galaxies and obtain number 
densities closer to the currently observed ones
in galaxy surveys (e.g. DESI, {\it Euclid}).

\subsection{Abundance-matching the full-physics and DM-only simulations}
\label{sec:sham}
We implement a SHAM prescription, where we rank-order
the subhalos in the DM-only run based on one of
8 proxy parameters
and those in the full-physics run based on 
their stellar mass. We do not input any scatter
into the relation, as we are interested in 
seeing whether the assumption of perfect 
correlation between the SHAM property and
stellar mass would yield good agreement
with the hydro simulation. Furthermore,
we want to compare overall trends
of the model using different parameters, so
not adding stochasticity makes this process
easier. The SHAM parameters 
we use as proxies for subhalo stellar mass are
\begin{itemize}
\item $V_{\rm relax}$: the maximum circular velocity reached 
by the subhalo during the periods of its history
in which it satisfies a relaxation criterion. The relaxation
criterion, the definition of which can be found in
\citet{2012MNRAS.427.1322L,2016MNRAS.460.3100C}, is 
motivated by the
assumption that a subhalo needs about one 
crossing time to return to equilibrium after
a major merger. Following \citet{2016MNRAS.460.3100C},
we approximate the crossing time at a given redshift 
as $t_{\rm cross} = 2 \ R_{\rm 200 m}/V_{\rm 200 m}
\approx 0.2/H(z)$ and compare it with the lookback
time to the moment where the subhalo reaches
3/4 of its mass at the considered redshift  
(the particular choice of 3/4 has been shown not to
make a difference to the final result). 
Finally, we conjoin all times where this
criterion is satisfied and find the highest circular
velocity of the subhalo across these epochs.
\item $V_{\rm peak}$: the maximum circular velocity reached 
by the subhalo throughout its history.
\item $V_{\rm max}$: the maximum circular velocity of the
subhalo at the final time (i.e. $z=0$).
\item $V_{\rm infall}$: the maximum circular velocity of the subhalo
at the time when it falls into a larger halo and 
ceases to be recognized as a central.
\item $M_{\rm peak}$: 
total mass of the bound particles in a subhalo
at the time when it achieves its peak mass.
\item $M_{\rm infall}$: 
the total mass of bound particles in a subhalo
at the time when the subhalo
infalls into a (larger) halo and becomes a
satellite of that halo. For central subhalos,
this quantity is equivalent to $M_{\rm SUBFIND}$,
\item $M_{\rm SUBFIND}$: the total mass of
bound particles in a subhalo defined by
the \textsc{SUBFIND} algorithm at the present time.
\item $M_{\rm circ,max}$: 
the total mass of the particles within the radius of
$V_{\rm max}$.
\end{itemize}

\begin{figure}
\centering  
\includegraphics[width=.5\textwidth]{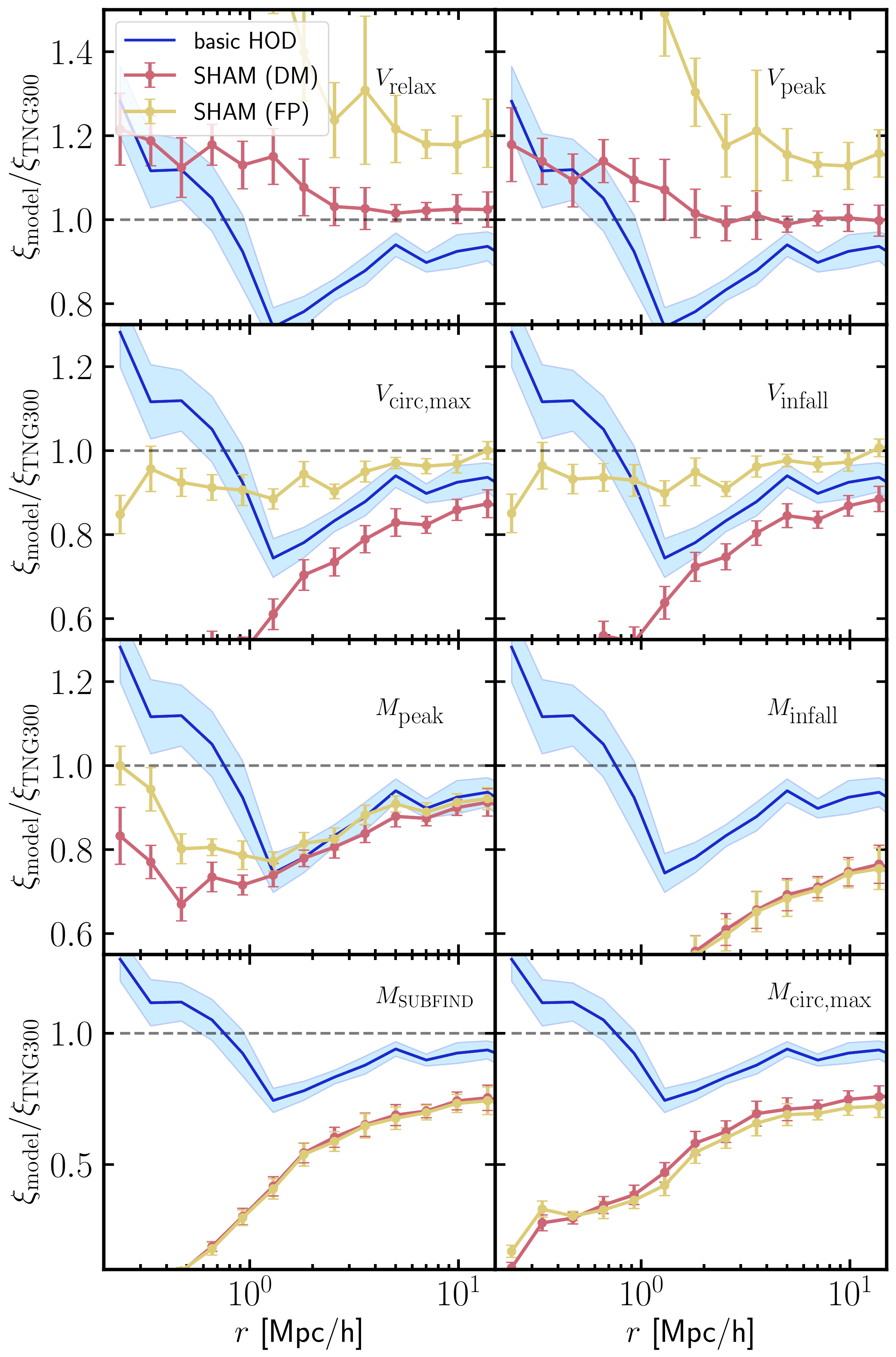}
\caption{Ratios between the auto-correlation of the
SHAM-model galaxies and the TNG300 galaxies,
using different luminosity proxies for the subhalo
abundance matching model (SHAM). 
The denominator of these ratios, i.e. the TNG300 sample, is
derived by selecting the most stellar-abundant subhalos
and computing their correlation function.
Here we compare this with two implementations
of SHAM: the first one (\textit{red solid line}) uses the standard
approach of rank-ordering the \textbf{DM-only}
subhalos in terms of one of the 8 SHAM properties
considered; in the second
case (\textit{gold solid line}), we have rank-ordered the
\textbf{full-physics} subhalos. The DM-only curves, SHAM (DM), suggest that
none of the SHAM parameters reproduces the observed
clustering of the stellar-mass-selected subhalos
perfectly on all scales, although $V_{\rm relax}$ and $V_{\rm peak}$
do substantially better than the rest of the parameters,
particularly the mass-based ones. In the alternate
SHAM model implementation, SHAM (FP), these parameters
overpredict the clustering.
In \textit{shaded blue}, we show the ``basic'' HOD
model for comparison, which exhibits a $\sim 12\%$ 
discrepancy on large scales.} 
\label{fig:sham}
\end{figure}

In Fig. \ref{fig:sham}, we illustrate the ratio 
between the two-point 
correlation function of the SHAM model
for the 8 parameter choices listed above
and the two-point correlation function 
of the TNG300 hydrodynamical galaxy sample.
We explore two different scenarios of applying
the SHAM model: in the first scenario, we
rank-order
the subhalos in the DM-only simulation by
some SHAM property and compute their
two-point correlation function, 
while in the second scenario, we order the subhalos
in the full-physics run and compute their
two-point statistics. In both scenarios,
we juxtapose these results with the 
auto-correlation of the full-physics subhalos
rank-ordered by stellar mass. We note that
typically one does
not have access to the hydro simulation
results, and so the results using
the DM-only subhalos are of greater relevance to future
mock catalog schemes.
We study both scenarios in an attempt to
understand the effects of baryonic physics
on the quantities used in empirical models
(see Section \ref{sec:bar} for a more extended 
discussion). As is the case throughout
this paper, we limit our analysis to the top
12000 subhalos in each rank-ordered list
so as to avoid resolution effects
and work with galaxy number densities 
comparable to those of current surveys
such as DESI and {\it Euclid}.

A striking observation
is that the clustering obtained using velocity-based
parameters exhibits a much smaller discrepancy (with respect to the full-physics simulation)
than when using the mass-based proxies for SHAM.
The parameter that yields the closest match is the
peak velocity, $V_{\rm peak}$, capturing the
clustering on scales larger than $\sim 2$ Mpc$/h$
with subpercent accuracy. On scales smaller than this, however,
the $V_{\rm peak}$-based SHAM model overpredicts the clustering by about 15\% 
\citep[see also][]{2020arXiv200503672C}.
Importantly, it deviates much less than the
``basic'' (mass-only) HOD model
when compared with TNG300. The next best performing SHAM proxy 
is the relaxation velocity, $V_{\rm relax}$, which
overpredicts the clustering only slightly (by about 3\%)
on large scales.
We have also experimented with introducing 0.125 dex 
scatter in the $V_{\rm peak}-M_{\rm star}$ relationship
and have found that it results in a marginal decrease
(of about $1-2$\%) in the clustering of the SHAM relative to 
TNG300, but it does not change the qualitative conclusions
from this analysis.

Fig.~\ref{fig:sham} makes it manifestly clear that
all of the mass-based SHAM proxies provide
a poor fit to the galaxy auto-correlation function
on all scales. The  suppression in the small-scale
clustering relative to TNG300 is particularly 
conspicuous, but even on large scales the
differences are as large as 30\%. The best performing mass-based
parameter is $M_{\rm peak}$. However, its performance
is still worse than that of the mass-only HOD sample.
Interestingly, applying SHAM to the full-physics subhalos
using $M_{\rm peak}$ does not boost the clustering
as much as it does for the velocity-based parameters.

Comparing the \textit{gold} and \textit{red}
curves in Fig.~\ref{fig:sham}, we see that the full-physics
SHAM objects are almost always more clustered
than in the ordinary case where the SHAM is applied to the DM-only simulation. The only parameters
for which this is not the case are the three mass
parameters $M_{\rm infall}$, $M_{\rm SUBFIND}$
and $M_{\rm circ,max}$. This finding, 
together with the observation that
for these three parameters the clustering is
substantially lower than TNG300, indicates
that there is no strong correlation between
the abundance of luminous components and 
these mass-based quantities and that the inclusion of 
baryonic physics does not play a
substantial role in improving their behavior. On the other hand,
we observe a significant upsurge in the clustering
for the remaining parameters. $V_{\rm relax}$ and
$V_{\rm peak}$ are especially interesting, where the full-physics SHAM
model yields higher auto-correlation power than
TNG300. This suggests that these
parameters are well-correlated with the subhalo
stellar mass and are therefore good SHAM proxies.
To completely reconcile the differences, one
could consider introducing scatter into the relationship.
The fact that the DM-only SHAM population has
a lower clustering indicates
that the removal of baryonic physics changes the
ranking of the subhalos and likely adds
stochasticity to the process. This is considered
in more detail in \ref{sec:bar}.

As a further test, 
we have applied the SHAM model to only those
DM-only subhalos that have matches in full physics and
compared their clustering to that of the top 12000 
full-physics subhalos that also have reliably found
counterparts (about 91\% of the 12000 most massive subhalos).
This results in a marginal improvement
in reconciling the clustering difference (which we do
not show in Fig. \ref{fig:sham}), decreasing the inconsistencies
by roughly $\sim 3$\% relative to the DM-only case. This is most likely the case
since isolating only the subhalos which have survived
violent disruption events mitigates some of the
baryonic physics effects, but not to a sufficient
degree. To fully understand the effect of
small-scale galaxy processes, a more thorough investigation
is needed.

\begin{figure*}
\centering  
\includegraphics[width=1.\textwidth]{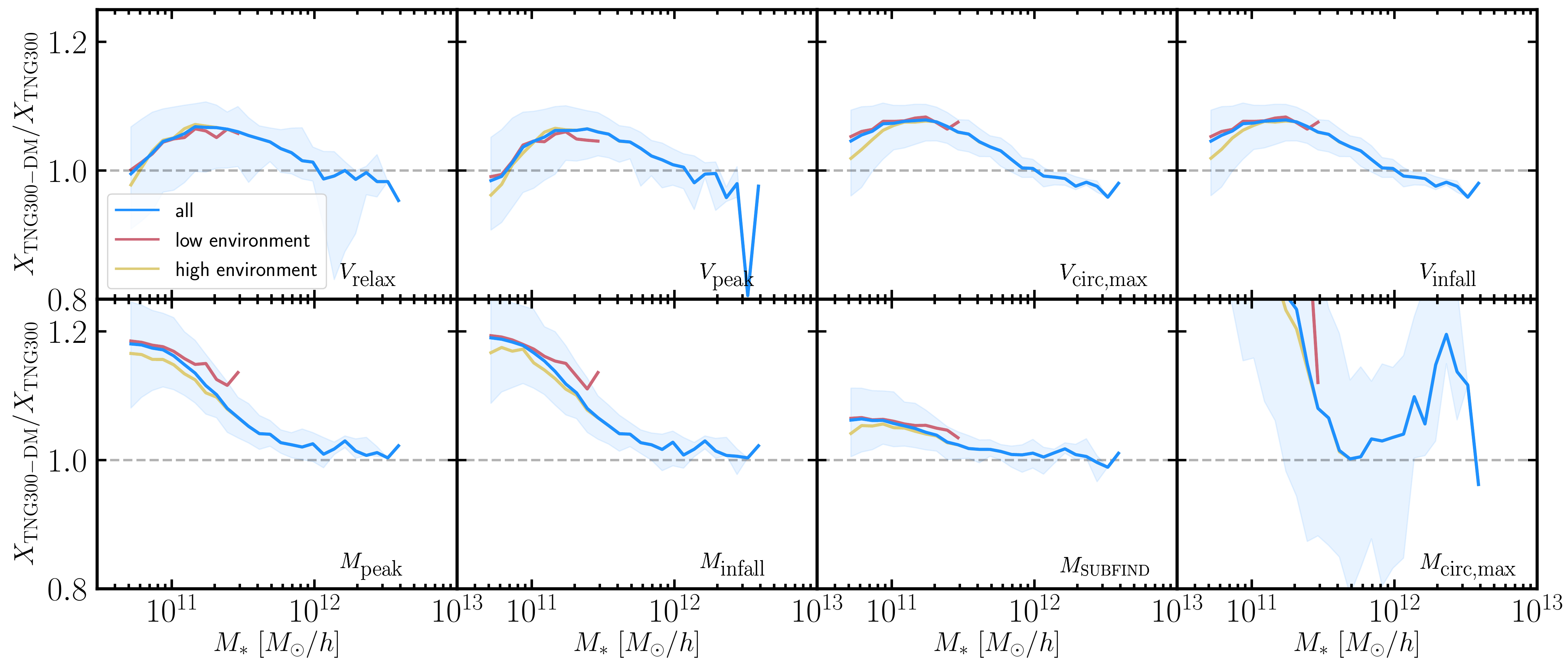}
\caption{Ratio between the 
different SHAM properties in the DM-only run
and their matches in 
the full-physics simulation: 
${X}_{\rm TNG300-DM}/{X}_{\rm TNG300}$, where
${X}$ denotes any one of the eight quantities
used as SHAM proxies (indicated in the 
bottom-right of each panel), and $\rm{TNG300}$ 
and $\rm{TNG300}$-${\rm DM}$ refer to the 
full-physics and
dark-matter-only simulations, respectively.
These show that the DM-only subhalos tend to
to have consistently higher values for the
various SHAM quantities when compared with
their full-physics counterparts. The most
substantially affected parameter is $M_{\rm 
circ,max}$, which provides a measure of the
mass distribution in the innermost parts of
the subhalo.
We also show the median curves for the subhalos
living in the densest (\textit{red solid lines}) 
and least dense (\textit{gold solid lines}) regions
of the simulation. These imply that subhalos in 
dense environments tend to retain their properties 
when one has switched on baryonic effects.}
\label{fig:sham_frac}
\end{figure*}

\subsection{Baryonic effects on the subhalos}
\label{sec:bar}
In this section, we delve into some detail to
try to understand
how the inclusion of baryonic physics affects
the subhalo quantities typically used in the
SHAM prescription. The SHAM approach
rests on the assumption that certain 
parameters are more tightly linked
to the amount of stellar material present in
a subhalo and rank-ordering them is
likely to recover
the galaxy population more faithfully. These parameters
are thought to be more resilient to 
well-known effects such as tidal
stripping, satellite disruption and 
gas expulsion (in the form of AGN feedback,
supernova feedback, stellar winds, etc.),
meaning that these quantities are less
likely to be dramatically changed by
galaxy formation physics. However, Fig. \ref{fig:sham}
suggests that the SHAM quantities of
interest (i.e. the ones based on velocity)
do change significantly between the
two simulation runs (TNG300 and TNG300-DM).
We estimate the impact of these
effects by tracking the same objects
in the two simulations and seeing how
their properties differ.

The comparison between the two simulations
is illustrated in Fig. \ref{fig:sham_frac}.
The median
curve and the 68-percentile contours are shown in {\it blue},
while in \textit{gold} and \textit{red},
we have selected the subhalos that live in
the densest and least dense environments
(divided into two equal halves),
respectively. In all cases, the effects
are largest ($10-20\%$) 
for the smaller subhalos,
which are most likely to be subjected to disk
disruption events and other violent processes.
We notice that the low-mass objects in the 
velocity-based panels display a downturn,
which somewhat diminishes the differences
between full-physics and DM-only. This effect
is likely part of the reason that the velocity-
based parameters perform overall better in
our SHAM modeling (see Fig.~\ref{fig:sham_frac}).
To illustrate, $V_{\rm peak}$ and
$V_{\rm relax}$ are good luminosity proxies,
as suggested by the \textit{gold} curves in the two top
panels of Fig. \ref{fig:sham}, and also
deviate less on the low-mass end than the other 6 parameters.

In order to explain the substantial difference in the two-point
correlation function between DM-only
SHAM and full-physics SHAM, it is helpful to realize that
the DM-only quantities have received 
a non-trivial amount of scatter with respect to
the full-physics ones in addition to the fact that they are substantially biased (as seen in Fig.~\ref{fig:sham_frac}).
Such stochasticity
inevitably results in a decrease of the clustering,
as this effective reordering of rank-ordered lists 
leads to assigning galaxies to subhalos that 
are in fact placed lower in the TNG300 lists.
From Fig. \ref{fig:sham_frac}, one can also
notice a modest dependence of the ratio
on the local environment of the subhalo.
For all parameters, the low-mass DM-only subhalos
living in dense environments (\textit{gold curves})
seem slightly more similar to their full-physics counterparts.

The effect of baryons on $M_{\rm circ,max}$, which measures the
mass in the inner core of the subhalo, is shown
in the last panel of Fig. \ref{fig:sham_frac}
to be particularly pronounced. The
noticeable difference between the hydro simulation
and the N-body simulation can be attributed
to energetic feedback events which expel energy and gas outwards from the center of the subhalo.
For even the larger
subhalos, the full-physics cores of
galaxies above $10^{12}$ $M_\odot/h$ 
are more devoid of matter. This is approximately the scale
where AGN feedback becomes important in driving out
gas from the inner parts of the subhalo.

Baryonic physics seems to have significant effects on our
ability  to match subhalos across the two simulation boxes. 
Out of the 12000 most massive galaxies in the full-physics 
simulation, we managed to find counterparts for only
91\% of them. At twice this galaxy number density,
the percentage of successful matches drops down by 
another 10\%. While in some cases their particles
have arbitrarily ended up in other subhalos, 
a substantial portion of them are in fact destroyed
by galactic disks in dense environments. This could
lead to a non-negligible bias in the one-halo term,
as the DM-only simulation produces more
subhalos that are eligible
to be populated through a SHAM algorithm
while in a ``true'' full-physics scenario, those would
have been much fewer.  

We have further studied the dependence 
of peak and infall statistics of satellites in
different environments as a function of subhalo mass.
The statistics we explored in the two simulations
were $\Delta M_{\rm peak|infall}$ and $z_{\rm peak|infall}$,
where $\Delta M_{\rm peak|infall}$ refers to the amount
of mass lost or gained when a subhalo reaches its maximum
circular velocity (``peak'') or when 
a subhalo first becomes
a satellite (``infall''). No significant
trends were observed apart from a larger scatter between
the full-physics and DM-only
redshift quantities of subhalos living in dense environments,
and we hope to further analyze this in the future.

\subsection{Dependence on galaxy properties}
\label{sec:split}
In this section, we explore how the SHAM
approach performs when applied to
subsamples of the galaxy populations.
We consider the same population of the 12000
most massive galaxies (corresponding to
a stellar mass cut of $M_{\rm star} \geq 5.1 \times 10^{10} M_\odot/h$)
used in the previous section
and split them into the following 5 pairs of
subgroups
\begin{itemize}
\item \textbf{low-mass/high-mass galaxies}:
we denote the subhalos with stellar masses below $8.3 \times 10^{10}
M_\odot/h$ as ``\textbf{low-mass}'' (6000 obj.), while
those above as ``\textbf{high-mass}'' (6000 obj.).
This choice is made so that the number of objects is 
the same in both samples. \\
\item \textbf{red/blue galaxies}: these are split by their $g-r$-band
colors. Subhalos with $g-r > 0.78$ correspond to the ``\textbf{red}''
population (6000 obj.) and subhalos whose $g-r$ value is below that threshold
are marked as ``\textbf{blue}'' (6000 obj.). The threshold is chosen so that
it roughly divides in two the bimodal color-mass distribution
and keeps the numbers equal. The
colors are obtained by a stellar population synthesis model and 
convolved an SDSS filter \citep[for details, see][]{TNGbim}.
\item \textbf{satellite/central galaxies}: we define the 
\textbf{``central''} (8934 obj.) to be
the largest subhalo within a parent halo,
while the rest of the subhalos are called
``\textbf{satellites}'' (3066 obj.).
Note that there is only one central in every halo.
\item \textbf{star-forming/quiescent galaxies}: similarly to
color, the threshold here is determined as the line in
sSFR-mass space (specific-star-formation-rate-mass space)
that approximately separates the two modes of the galaxy population
into ``\textbf{star-forming}'' ones (4800 obj.) and ``\textbf{quiescents}'' (7200 obj.), and equals
$\log ({\rm sSFR}) = -13.86$,
where the specific star formation rate is measured in units
of $\rm{yr}^{-1}$.
\item \textbf{early-forming/late-forming galaxies}: 
these are defined by their ratio between the mass
at present time (i.e., $z=0$) and their
mass at $z = 0.3$; we denote this ratio as $f_{\rm growth}$.
Subhalos for which the mass gain
in the past 3.5 billion years is larger than
the median ($f_{\rm growth} = 0.86$) are said to be
``\textbf{early-forming}'' (6000 obj.), while galaxies
for which the growth is smaller are ``\textbf{late-forming}''
(6000 obj.). The threshold is selected so that the 
number of objects in the two groups is roughly equal.
The growth factor, $f_{\rm growth}$, can be defined self-consistently
for each subhalo and thus avoids comparisons of the
galaxy populations
at different redshifts, which might be subject to
resolution effects of the simulation or the merger
tree. We have also tested other common definitions such
as the redshift at which the subhalo reached half
of its present mass and
the redshift at which it attained its peak velocity,
but none of them led to a substantial change in the
qualitative interpretation.
\end{itemize}
We have also checked that regardless of how the splits
are made and what number density we choose to work with, the 
qualitative conclusions remain unchanged.

Once we have applied the standard SHAM methodology
and obtained the two lists of subhalos (12000 in each):
one for the full-physics subhalos rank-ordered
by their stellar masses and one for the 
dark-matter-only subhalos rank-ordered by 
$V_{\rm relax}$, we further split the full-physics galaxies
into the subcategories defined above and
take the dark-matter-only subhalos
that have the same indices
in their rank-ordered list as the indices of the
full-physics objects in the selected subcategory. 
In this way, we investigate the
question of which galaxy sub-populations are
captured best after applying a simple SHAM prescription. 
We choose the SHAM property $V_{\rm peak}$,
the highest circular velocity a subhalo
attained throughout its history,
since this is the best-performing
SHAM parameter as seen in Fig. \ref{fig:sham}.

\begin{figure*}
\centering  
\includegraphics[width=1\textwidth]{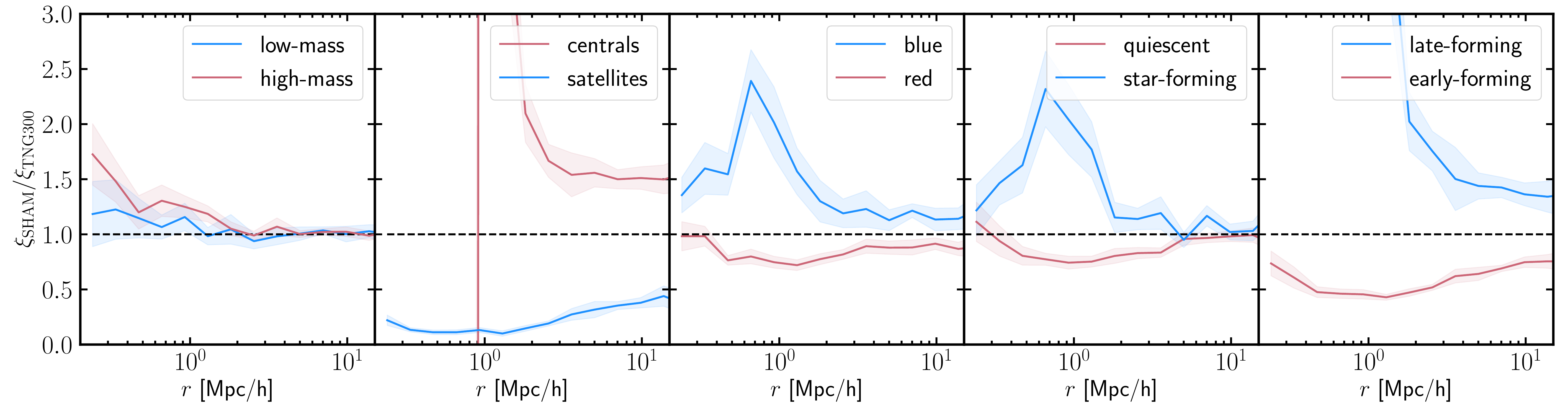}
\caption{Correlation function ratios of SHAM-model galaxies split into different populations
(defined in Section \ref{sec:split}), using
$V_{\rm peak}$ as the SHAM parameter. In the
\textit{leftmost panel}, we show the galaxy
population split in terms of stellar mass. \textit{First panel}:
We see that the low-mass galaxies two-point function is
captured better than that of the low-mass sample. \textit{Second panel}: The clustering of the
centrals is overpredicted by SHAM, while that
of the satellites is significantly lower.
\textit{Third and fourth panel}: These reveal
that the two-point statistics of blue and
star-forming galaxies are higher than that of the 
rank-ordered dark-matter subhalos. 
\textit{Fifth panel}: The correlation function
of
late-forming galaxies is significantly higher than that of their corresponding DM-only subhalos,
while that of the early-forming objects is
noticeably lower. The percentage of central galaxies
in the late-forming sample is $\sim 95.9\%$, which explains
the visual similarity between the \textit{fifth} and the \textit{second} panel.} 
\label{fig:sham_split}
\end{figure*}

We compute clustering statistics via the two-point correlation
function for the full-physics sub-populations
and for the corresponding DM-only subset of 
selected subhalos. In Fig. 
\ref{fig:sham_split}, we show the ratio between the
correlation functions for the 5 pairs of galaxy sub-populations
listed. The first panel suggests that the 
clustering of the high-mass 
galaxies is much better-captured by the SHAM approach
on small scales despite still exhibiting a discrepancy of 
about 20\% on small scales.
On the other hand, the low-mass galaxy clustering
deviates significantly from the abundance-matched
DM-only subhalos,
differing by about 30-60\% on small scales. 
The percentage of central galaxies in the high-mass
and low-mass subgroups are 79.1\% and 69.8\%, respectively,
which as we will see next plays a significant role
in predicting the clustering.

As can be seen in the second panel, the galaxies identified
as centrals tend to be more clustered by about 50\% on
large scales than their
corresponding DM-only subhalos in the SHAM catalog. 
This implies that particular care
should be taken when modeling the satellite population,
which is discrepant by at least 50-80\% on all scales. This result
provides a possible explanation for why the high-mass galaxies
in the {first panel},
which are predominantly centrals, are better modeled by SHAM.
The dark-matter properties of central galaxies are perhaps
more strongly correlated with their stellar masses, as
assembly bias and baryonic physics affect satellites
more prominently than centrals. This is because centrals
live in the cores of halos
and are believed to be shielded from environmental and
tidal stripping effects
while also being exposed to more violent merger events.

In the next two panels, we show the results for 
blue/red and star-forming/quiescent galaxies. The blue
and star-forming sub-populations, which have a significant
overlap, are substantially more clustered than their SHAM
counterparts (by about $15-20$\% on large scales and more 
than 50\% on small scales). In contrast,
the red and quiescent sub-populations underpredict the clustering,
although the observed differences appear to be
reconciled on large scales to within $5-10$\%.
This finding may be somewhat counter-intuitive,
as we often think of the red and quiescent galaxies as
being more massive and might therefore expect to see
the opposite effect. However, studying the central and
satellite fractions reveals that a larger percentage of
the blue and star-forming galaxies are centrals
(76.8\% and 80.9\%, respectively) than
their red and quiescent counterparts
(72.1\% and 70.2\%, respectively), which
combined with the result in the second panel
offers insight into why the bluer objects have stronger
clustering, at least within the sample selection used here.

The final panel shows the division into early-
and late-forming galaxies. It suggests a strong
correlation between the large-scale
distribution of the objects in a SHAM catalog obtained by
conditioning on $V_{\rm peak}$ and the late-forming
galaxies. One possible interpretation is that 
galaxies that vigorously accrete more mass relative to the remainder of the population 
in the last 3.5 billion years since $z = 0.3$ have had 
less opportunity to be stripped of their dark matter
envelopes, so for them, a dark-matter property 
tightly coupled to their assembly history such as 
$V_{\rm peak}$ remains more strongly correlated with 
the mass of the luminous component in full-physics.
Intriguingly, out of the late-forming galaxies,
about 95.9\% are centrals (and out of the early-forming
ones 53.0\% are centrals), which provides support for
the above claim that the late-forming galaxies are
less exposed to stripping and/or recent disruption activity.

\section{HOD modeling}
In this section, we describe a framework for developing
improved mock catalogs in preparation of future surveys.
To alleviate the tension between the ``basic'' HOD
model and the hydrodynamical simulation, we 
augment the HOD model with a secondary parameter in
addition to mass. The parameters included in the extended
HOD model are local environment, velocity anisotropy, 
virialized mass ($\sigma^2 R_{\rm halfmass}$), and
total potential energy. This two-dimensional HOD model
(2D-HOD) has one free parameter, $r$, which measures the
strength of the correlation between halo occupancy and the
secondary parameter within each 5\% mass bin and is
tuned to match the two-point galaxy correlation function
of the 2D-HOD galaxies with that of TNG300. Here, we
also explore alternative statistical tools to
help us select the best population strategy.

To ensure that the galaxy sample from
IllustrisTNG is robust,
we define our galaxies as subhalos with at least
10,000 gravitationally bound
star particles, which results in
a galaxy sample with a number 
density of $n_{\rm gal} \approx 1.3 \times 10^{-3}
\ [{\rm Mpc}/h]^{-3}$.
We have further checked the impact of
adopting different halo finders, working with a limited 
simulation volume, and incorporating cosmic variance effects
\citep[for details, see][]{2020MNRAS.493.5506H}.

\subsection{Parameters}
\label{sec:pars}
Here we review some of the parameter choices introduced in
\citet{2020MNRAS.493.5506H}, which result in the most substantial shifts
of the 2D-HOD galaxy auto-correlation function. In addition,
we revise the definition of ``local environment''
in order to make it less definition-dependent. We introduce
another parameter which was also found to be strongly
correlated with the galaxy clustering on large scales: the
total halo potential energy.

\subsubsection{Velocity anisotropy}
The ``mass-anisotropy degeneracy''
introduced by the Jeans model \citep{1987ApJ...313..121M}
predicts a strong relationship between the
mass profile of a distribution of particles 
and the velocity anisotropy of the orbits 
that trace the resulting potential,
Following \citet{1987gady.book.....B}, we define the velocity anisotropy as
\begin{equation}
    \beta = 1-\frac{\sigma_{\rm tan}^2}{2 \sigma_{\rm rad}^2},
\end{equation}
where $\sigma_{\rm tan}$ and $\sigma_{\rm rad}$
are the tangential and radial velocity dispersions, respectively.
We calculate these quantities over all particles in the
FoF halo by projecting the velocity of each particle
along and perpendicular to the radial direction
(defined with respect to the position of the particle with
the minimum gravitational potential energy) and then computing the
standard deviation of each component 
\citep{2019arXiv190302007R}.

Thus defined, $\beta$
depends on the shape of the halo and for that reason captures
information from the full phase-space structure of the parent halo. 
The limits of this parameter, $-\infty$ and 1, respectively, correspond to
radially and tangentially dominated velocity dispersions, while $\beta = 0$ indicates an isotropic
distribution of particle orbits.

We conjecture that a plausible explanation for
the relationship between a low value of $\beta$
and a high galaxy clustering, found in \citet{2020MNRAS.493.5506H}, is that halos which
have undergone recent accretion events tend to
have particles which exhibit higher tangential
velocities ($\sigma_{\rm tan}$) due to
deflections caused by gravity 
shortly before accretion.
This is particularly important for regions of
high density, where mergers dominate the mass growth of halos
\citep{2009MNRAS.394.1825F,2010MNRAS.401.2245F},
and are therefore presumed to be influential in determining the velocity 
structure of these halos and the richness of substructure 
that exists within them. 
As for low-density regions,
accretion occurs oftentimes gradually in the radial direction since the 
gravitational field is dominated
by the central galaxy. This
results in high values of the particle velocities in the radial direction, and so the 
value of $\beta$ increases
\citep{2009MNRAS.394.1825F,2010ApJ...708..469F}.

\subsubsection{Mass measure assuming
virial theorem}
One of the most widely accepted
choices for a virialized mass proxy is
$M_{\rm 200m}$, but there are many others 
such as those resting on the assumptions
of the virial theorem
\begin{equation}
    \frac{G M_{\rm vir}}{R_{\rm vir}} = \sigma^2 ,
\end{equation}
where $\sigma$ is the dispersion velocity of the
particles in the halo. Here we consider
the combination $\sigma^2 R_{\rm halfmass}$,
where $\sigma$ is the velocity dispersion of the most
massive subhalo for a given halo,
while $R_{\rm halfmass}$ is the halfmass 
radius  of the most massive subhalo. 

A possible explanation for why this quantity is tied to
the number of galaxies hosted by a halo is that
$\sigma^2 R_{\rm halfmass}$ encodes a dynamical 
description of the halo merger history and is 
related to halo concentration (i.e. its central density).
Its relationship with concentration is two-fold: on one hand,
the dispersion velocity of the central subhalo
is a dynamical proxy for the subhalo mass 
and the halo concentration,
as more concentrated halos are expected to have 
relatively higher dispersion velocities of their main subhalos,
and on the other, central
subhalos with large $R_{\rm halfmass}$ 
are more likely to have consumed
the smaller subhalos surrounding them resulting
in higher central density. These two effects imply
that there are more satellite
galaxies on average for an object with a 
small value of $\sigma^2 R_{\rm halfmass}$.

\begin{figure*}
\centering  
\includegraphics[width=1\textwidth]{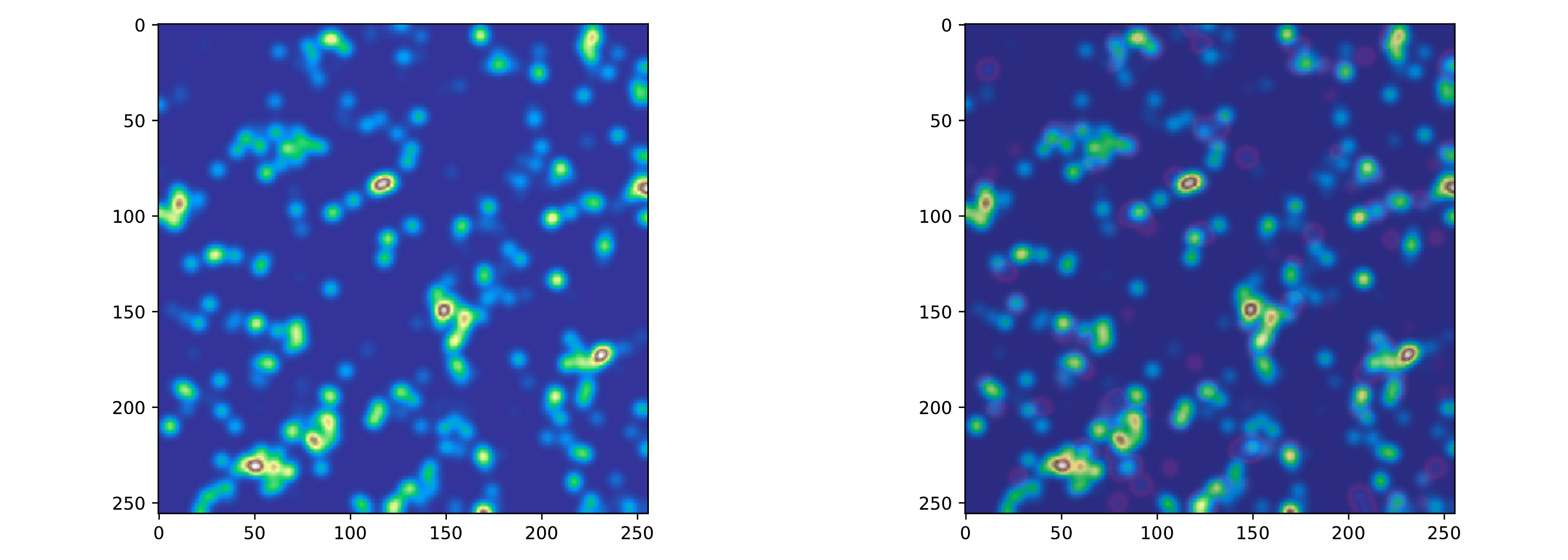}
\caption{Visual motivation for augmenting the HOD 
model with an environment parameter.
\textit{Left panel:} the smoothed density field
of the ``true'' galaxies in the TNG300 
hydro simulation. \textit{Right panel:} an overlay
between the density field of the ``true'' galaxies 
(same as left panel) and the difference between
that and the density field of the ``basic''
HOD model (in red). The red circles surrounding
many of the lower-density regions indicate an excess of galaxies
around smaller clusters with respect to the 
``true'' population. These are not observed around
the densest clusters, suggesting that the ``basic'' (mass-only)
HOD model fails to supply a sufficient number
of galaxies in these regions. The ``basic'' HOD model
used here
preserves the total galaxy number in the simulation,
and in this figure we only show the positive
difference between the two density fields, i.e.
$\Delta \delta > 0$, where
$\Delta \delta \equiv \delta_{\rm bHOD} - \delta_{\rm TNG300}$.}
\label{fig:dens_shuff}
\end{figure*}

\subsubsection{Local environment}
A halo residing in a dense region is expected to contain
more galaxies on average than a halo in an underdense
region. This is because halos in overdense regions 
experience more mergers, whereas those in underdense 
regions have more mass accreted in the form of smooth material
\citep{2007MNRAS.378..641A,2017A&A...598A.103P,
2018MNRAS.476.5442P,2018MNRAS.473.2486S}.
We assess the effects of local environment of halos on 
the large-scale clustering, adopting the following
definition:
\begin{itemize}
\item[1.] We evaluate the density field,
$\delta (\mathbf{x})$, using cloud-in-cell (CIC) interpolation on a 
$256^3$ cubic lattice of the DM particles. Each
cell is of size
$205/256 \ {\rm Mpc}/h \approx 0.8 \ {\rm Mpc}/h$.
\item[2.] We smooth the density field with a Gaussian
kernel of smoothing scale
$R_{\rm smooth} = 1.1 \ {\rm Mpc}/h$.
\item[3.] The local environment parameter is determined
by the value of the smoothed density
field in the cell its center-of-potential is located in.
\end{itemize}
Conditioning on this parameter leads to a substantial
increase in the galaxy clustering on large scales.
In Fig. \ref{fig:dens_shuff}, we provide a visual motivation
for using environment as a secondary parameter in a
mass-only HOD approach. Each panel illustrates
a slice of the smoothed galaxy density field split into
$256^3$ cells with a smoothing scale of $R = 3
\ {\rm Mpc}/h$ of the ``true'' galaxy distribution
(\textit{left panel}), with the \textit{right panel} 
painting on top the difference between the
two density fields, $\Delta \delta = \delta_{\rm bHOD} -
\delta_{\rm TNG300}$. Here we only present the positive
values of this difference ($\Delta \delta > 0$)
so as to show clearly the excess
of galaxies, denoted with red circles,
that the ``basic'' (mass-only) HOD model predicts with respect to TNG300.
On the other hand, because the total occupation of the
halos is preserved in the ``basic'' HOD model (see Section
\ref{sec:bhod}), the lack of red circles around the densest
clusters signifies that this model fails to populate
the denser clusters with enough galaxies. For this reason,
considering a secondary property which preferentially populates
halos living in denser environments ought to resolve
some of the tension with the hydrodynamical simulation.

\subsubsection{Potential energy}
The gravitational potential energy of each particle
is evaluated as follows
\begin{equation}
    \Phi(\mathbf{x}_j) = - G \sum_{i=1,N}^{i \neq j} m_i
    g(|\mathbf{x}_i-\mathbf{x}_j|)
\end{equation}
where $g(r)$ is the particle pair potential evaluated by \textsc{AREPO}.
Since $\Phi(\mathbf{x}_i)$ is obtained as a summation
over the entire box, it is closely related to the
large-scale environmental properties around a given
particle and thus serves an excellent proxy for the cosmological environment.

We explore two halo parameters derived from the particle
potential energy: total potential energy and minimum
potential energy. The former is defined as the sum of
the gravitational potential of all particles within
twice the virial radius ($=R_{\rm 200 m}$) of the halo,
while the latter is simply the smallest value of the
potential energy across all particles within the halo.
We have checked that the choice of where the summation
for the total potential energy stops makes a negligible
difference to the subsequent analysis. Furthermore, the
results obtained by using the minimum potential energy
are also very similar to those using the total potential
energy. Hence, we will hereafter concentrate exclusively
on the total potential energy parameter, but the 
conclusions drawn apply to our other measures of halo
potential energy as well.

\subsection{Statistical comparisons with the TNG300 galaxy sample}
In this section, we consider alternative measures
of the large-scale galaxy distribution such as 
cross-correlation functions, void size distributions and
cumulants of the smoothed galaxy density field to compare
the ``fitted'' 2D-HOD galaxy samples with the ``true'' TNG300
galaxy sample.

\subsubsection{Bias and correlation coefficient}
\label{sec:gm}
Most of the cosmological information of the matter 
distribution is encapsulated
in the power spectrum (or correlation function)
of the matter density fluctuations as a function
of scale and redshift. 
However, galaxies are not perfect tracers of the
underlying mass distribution, and thus, it is 
important to understand the relationship between the 
large-scale distribution of matter and that of galaxies. 
The galaxy auto-correlation function is related to the
matter correlation function, $\xi_{mm}(r)$, through
the real-space galaxy bias, $\tilde b$,
in the following way
\begin{equation}
    \xi_{gg}(r) = \tilde b^2(r) \xi_{mm}(r).
\end{equation}
One of the most popular current methods for estimating
the bias is through the
galaxy-matter cross-correlation function,
$\xi_{gm}(r)$, which can related to the matter 
two-point correlation function 
through $\tilde b$
and the real-space cross-correlation 
coefficient between matter and galaxy fluctuations, 
$\tilde r$ \citep{2008MNRAS.388....2H,2018PhR...733....1D}:
\begin{equation}
    \xi_{gm}(r) = \tilde b(r) \tilde r(r) \xi_{mm}(r)
\end{equation}
where the galaxy bias is
\begin{equation}
    \tilde b(r) = \Bigg[\frac{\xi_{gg}(r)}{\xi_{mm}(r)}\Bigg]^{\frac{1}{2}}
\end{equation} 
and the correlation coefficient is
\begin{equation}
    \tilde r(r) = \frac{\xi_{gm}(r)}{[\xi_{gg}(r) * \xi_{mm}(r)]^{1/2}} .
\end{equation} 

\begin{figure*}
\centering  
\includegraphics[width=1\textwidth]{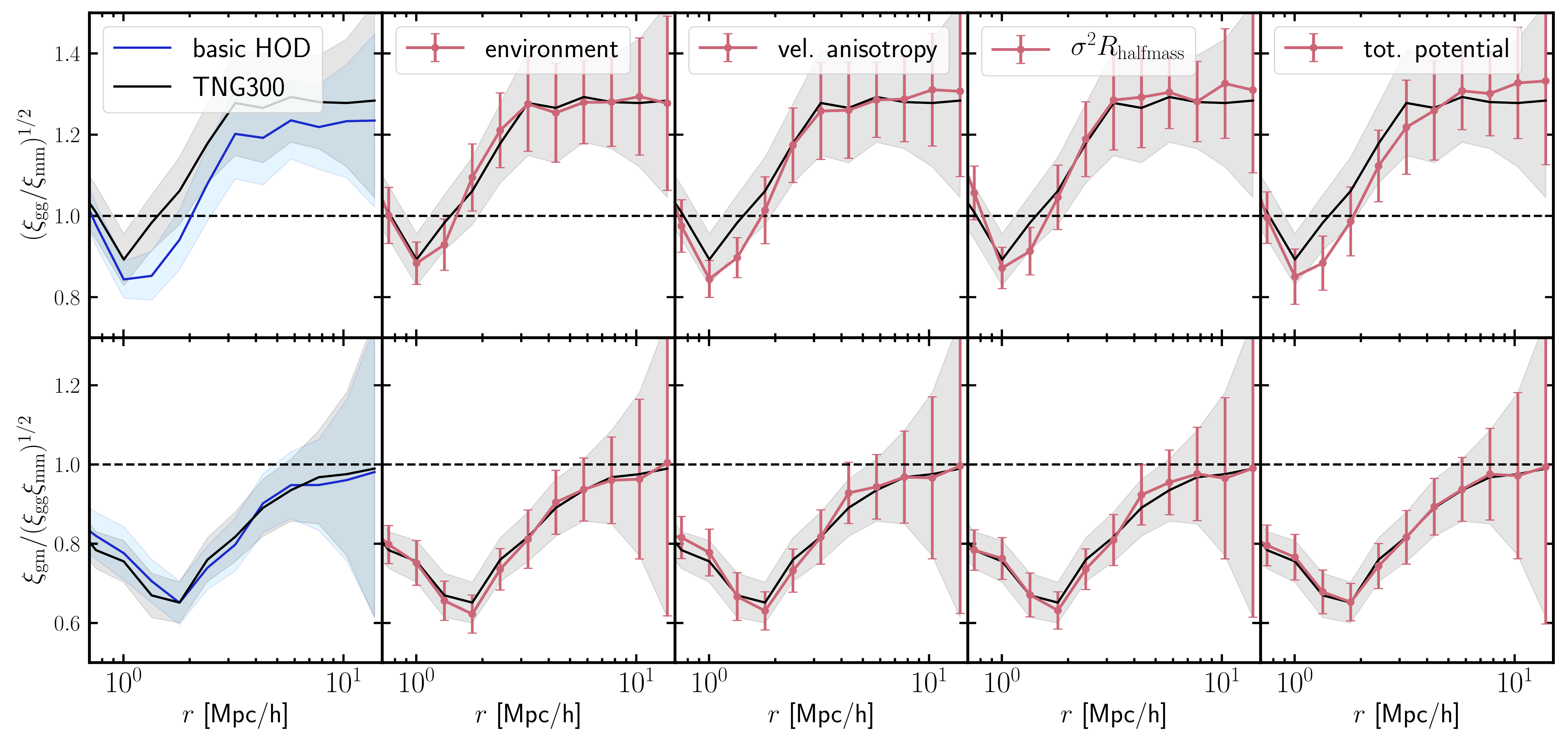}
\caption{Bias and correlation coefficient of
the ``true'' galaxies in TNG300
and those obtained for the HOD prescriptions
considered in this paper. \textit{Top
panels}: these show the galaxy bias ($\tilde b(r)$)
defined as the 
square root of the ratio of the galaxy and
matter auto-correlation functions, while the 
\textit{bottom panels} show the real-space 
correlation coefficient, $\tilde r(r)$. In shaded
blue we show the curves for the TNG300 (``true'') 
galaxy population, while in orange
we show the ``basic'' (mass-only) HOD samples
as well as the ``fitted'' 2D-HOD ones. The
galaxy bias goes to a constant on large scales,
and the correlation coefficient approaches 1,
suggesting that a linear bias approximation on
scales beyond 10 Mpc$/h$ is appropriate. We see
that the agreement between all models
is excellent for the correlation coefficient,
while in the case of galaxy bias, the differences
are as expected from the two-point correlation
statistics -- namely, the ``basic'' HOD model
has lower clustering than the ``true'' sample,
while the 2D-HOD models show a good agreement with
the hydro simulation, as they have been designed to
fit it. The relatively larger size of the error
bars compared with the ratio plots (shown in the
rest of the paper) is due to a cancellation effect 
present when taking ratios of quantities.} 
\label{fig:bias_corr}
\end{figure*}

In Fig. \ref{fig:bias_corr}, we demonstrate what these look
like for the baseline HOD galaxy distribution
and the 2D-HOD galaxy samples compared with the 
TNG300 (``true'') galaxy sample.
We see that on large scales (larger than a few times the typical size of 
a dark matter halo), the galaxy bias tends to a
constant value, so we can use the linear bias
approximation to infer the underlying matter distribution
\citep{1980lssu.book.....P,1996MNRAS.282..347M,2013MNRAS.432.1544M}.
For the linear bias approximation to be valid,
the cross-correlation coefficient is also expected to be 
scale-independent on large scales, approaching unity
\citep{2010PhRvD..81f3531B}.
As long as one considers large-scale galaxy
clustering on scales
much greater than Mpc scales, the observed correlation
should be sourced from the gravity field of the
total matter. 

In this large-scale regime, we notice that 
discrepancies between the galaxy auto-correlation functions
of TNG300 and any galaxy population model manifest
themselves in the galaxy-matter correlation function
at approximately \textit{half} their level, i.e.
$\xi_{gm}(r)$ differs from the ``true'' galaxy
population on large scales half as much as $\xi_{gg}(r)$
does because no physics besides gravity can change the
large-scale bias. In addition, our simulations employ
adiabatic initial conditions, so there is only one
degree-of-freedom on large scales, i.e. the matter
distribution. 

One can explain this factor-of-2 rule in the following way.
Let the ``true'' galaxy distribution
in the TNG300 hydro simulation
sample be denoted as $g$, and that of our alternative
population model be $\tilde g$. Since on large scales,
the galaxy bias can be well-approximated as linear,
we can express the latter distribution as $\tilde g = g(1-\epsilon)$ where $\epsilon$ is a small number.  
The galaxy-matter power spectrum is then $\xi_{\tilde gm}(r)
\propto (1-\epsilon) \, \xi_{gm}(r)$, while 
the auto-correlation function is $\xi_{\tilde g \tilde g}
\propto (1-\epsilon)^2 \, \xi_{gg} \approx (1-2 \epsilon) \,
\xi_{gg}$.
So a discrepancy of $\sim$ 12\% between the 
auto-correlation of the ``basic''
HOD model and the ``true'' galaxy distribution
\citep{2020MNRAS.493.5506H} is expected to manifest
itself as a $\sim \! 1/2 \times 12\% = 6\%$ difference in 
the galaxy lensing probe. 

However, it is important to note that while on large scales
the linear bias approximation appears to be viable, it certainly
breaks down on smaller scales ($\sim 1$ Mpc$/h$). This 
has important implications for
analyses using mock catalogs created via phenomenological approaches
such as the HOD framework. The small-scale signal encodes a lot
of information about cosmological parameters such as
$\Omega_m$ and $\sigma_8$. In addition, modeling these scales correctly is a key
requirement for shear analysis. Finally, the small-scale
data provide an important window for probing different DM models and 
understanding the effects of baryonic physics.
It is reassuring to see that the galaxy bias is in a 
reasonable agreement between the
different 2D-HOD models and that
the discrepancy known in the two-point
correlation statistic manifests itself as expected on the galaxy
bias measurement using the ``basic'' HOD model.
The total potential 2D-HOD distribution exhibits
the worst agreement with the ``true'' galaxy bias out of all the parameters considered --
not only is it discrepant from the TNG300 galaxies, but
it also shows a scale-dependence which violates
the assumption of constant linear bias on large scales.
Furthermore, the cross-correlation coefficients 
derived for all models shown in Fig. \ref{fig:bias_corr}
exhibit a very similar behavior across all scales,
suggesting that the galaxy-matter cross correlation relates
similarly to the galaxy and matter clustering regardless
of the underlying population model. The minimum point of
the cross-correlation coefficient is on the outskirts of
the halo ($r\sim 2$ Mpc$/h$), between the one- and two-halo 
terms, where the dark matter outweighs
the luminous component, which has sunk to
the halo center due to dynamical friction.

\subsubsection{Galaxy-galaxy lensing}
\begin{figure*}
\centering
\includegraphics[width=1\textwidth]{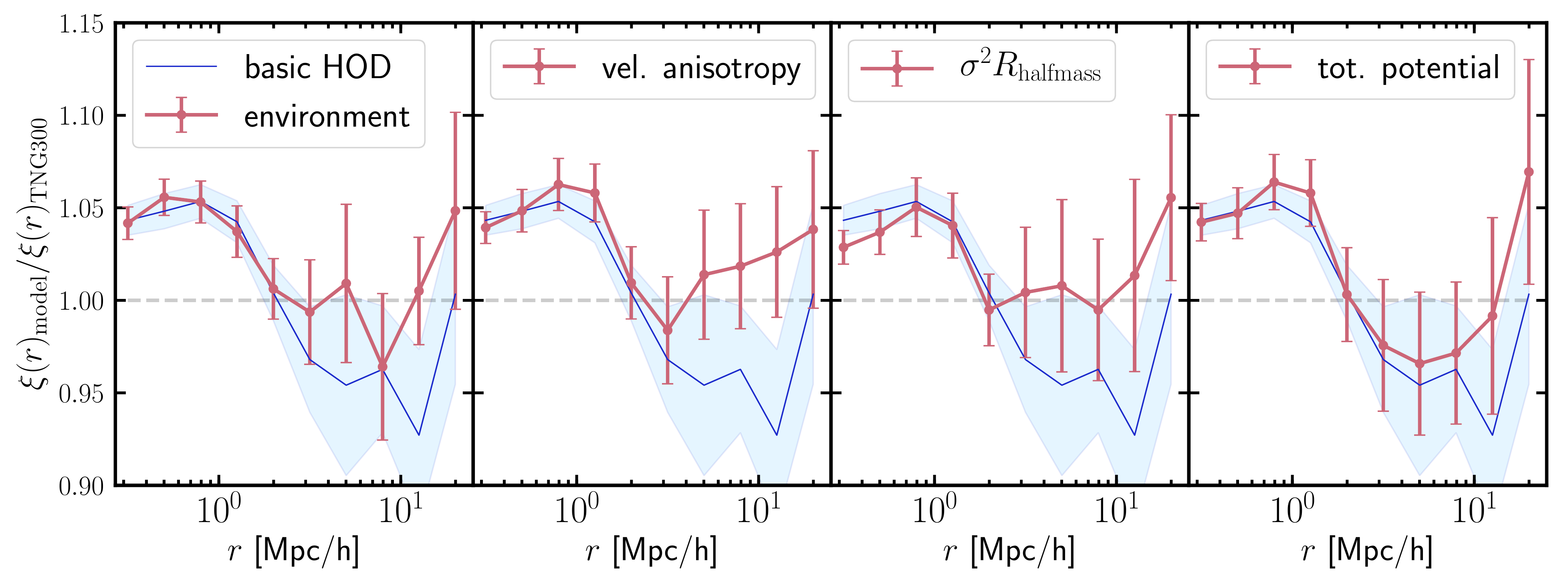}
\caption{Ratio of the galaxy-galaxy lensing 
excess surface mass density
between the ``true'' galaxies in TNG300
and those in the other HOD prescriptions
considered in this paper. The shaded blue curve
corresponds to the ``basic'' (mass-only) HOD
sample, while the orange ones are obtained using
the various choices of secondary parameters,
tuned to match the two-point correlation function
on large scales. We see that the mass-only
HOD model differs noticeably on large scales
by about 4\%, while the ``fitted'' galaxy samples
agree better with the TNG300 distribution.
The total potential sample and the $\sigma^2
R_{\rm halfmass}$ seem to exhibit the largest
amount of discrepancy.}
\label{fig:gal_lens}
\end{figure*}

The cross-correlation of large-scale structure tracers with
the shapes of background galaxies, referred to as stacked
lensing or 
galaxy-galaxy lensing, offers a unique statistical method for measuring 
the average total matter distribution around foreground objects.
Stacked lensing measurements are expected to be one of the most powerful 
probes for ongoing and upcoming galaxy surveys, allowing cosmologists
to address fundamental physics questions such as the nature of dark energy
and neutrino mass \citep[e.g.][]{2011PhRvD..83b3008O}.
Furthermore, by combining stacked lensing and auto-correlation measures 
of the same foreground galaxies, one can constrain cosmology 
by breaking degeneracies between galaxy bias and cosmological
parameters \citep[e.g.][]{2005PhRvD..71d3511S,2020arXiv200203867S}.

As a measure of the stacked lensing, here we 
consider the excess surface mass density profile, 
denoted as $\Delta\Sigma$. It is obtained by first calculating
\begin{equation}
\Sigma(r_p) = \bar{\rho}\int_0^{\pi_{\rm max}} \left[1+\xi_{\rm gm}(\sqrt{r_p^2+\pi^2}) \right]\mathrm{d}\pi,
\end{equation}
where $\bar \rho$ is the mean matter density while 
$r_p$ and $\pi$ are the distances perpendicular 
and parallel to the line of sight, respectively. Then
one can find the excess surface mass density as 
\begin{equation}
\Delta\Sigma(r_p) = \bar{\Sigma}(<r_p) - \Sigma(r_p),
\end{equation}    
where the mean surface mass density interior
to the projected radius is given by
\begin{equation}
\bar{\Sigma}(<r_p) = \frac{1}{\pi r_p^2}\int_0^{r_p}\Sigma(r_p^{\prime})2\pi r_p^{\prime} \mathrm{d}r_p^{\prime}.
\end{equation}

In Fig. \ref{fig:gal_lens}, we demonstrate the excess
surface mass density, $\Delta \Sigma$, for the various 2D-HOD models
proposed in this work and compare them with the ``basic'' (mass-only)
HOD in terms of their ratio with TNG300. We see that the mass-only
HOD model is at a 5\% tension compared with the hydro simulation
on large scales, whereas the four 2D-HOD galaxy distribution
proposals do a significantly better job at recovering the
``true'' galaxy statistic. The 5\% discrepancy is approximately
equal to half of the difference in the galaxy auto-correlation,
validating the naive calculation outlined in Section \ref{sec:gm}.
For the other four parameters, we see that all curves are in
relatively good agreement with TNG300, the most prominent misfit
being the total potential parameter, similarly to the conclusion
drawn from Fig. \ref{fig:bias_corr}. 

These two findings
together are a cause of concern for using the total potential
as a secondary HOD parameter, as they suggest that 
this parameter fails to mimic the large-scale behavior
of the galaxy clustering (i.e. the linear bias approximation)
as well as its relationship with the total matter distribution
at the required subpercent accuracy. The second worst
2D-HOD parameter is $\sigma^2 R_{\rm halfmass}$, which on large scales
is offset from TNG300 at $>1\sigma$. However, due to the
volume limitations of TNG300, we cannot place significant importance
on this observation by itself until we perform complementary tests.

\subsubsection{Cumulants of the density field}
Density field cumulants correspond to a 
set of statistics derived from measurements of the moments of 
the smoothed density field.
They can be understood as degenerate $N$-point
correlation functions or integrated monopole
moments of the bispectrum, which are closely 
related to neighbor counts both in their
physical interpretation as well as in their
algorithmic implementation. In this work, we are
interested in exploring alternative statistical tools to the two-point
correlation function for quantifying and
describing galaxy populations obtained from our
extended 2D-HOD model. Of particular
importance is the scale of galaxy clusters where we expect
that galaxy population methods could exhibit substantial
differences \citep{1994A&A...291..697B,1994MNRAS.268..913G}. For this reason, we explore the regime
between 3 Mpc/$h$ and 8 Mpc/$h$ in the subsequent analysis.

\begin{figure}
\centering  
\includegraphics[width=0.5\textwidth]{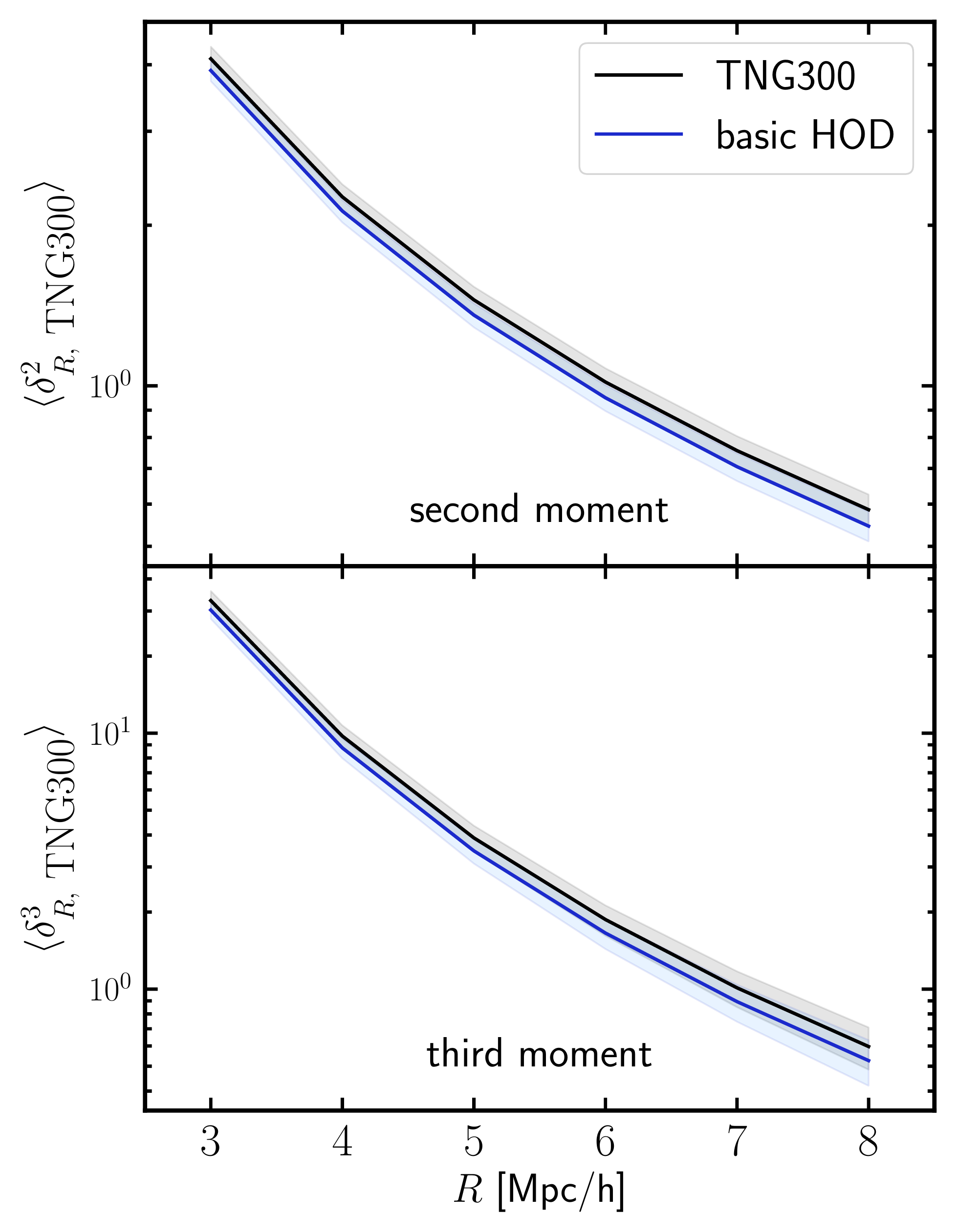}
\caption{Second and third moment of the smoothed
galaxy density field of the ``true'' TNG300 sample
at different smoothing scales ($3 - 8$ Mpc$/h$).
The errorbars are obtained by jack-knifing the 
final density distribution before measuring the
second and third moments.
As the density field gets smoother, the values
of the moments rapidly decrease. The discrepancy
between the two curves is illustrated more clearly
in the ratio plots in Fig. \ref{fig:2nd_3rd}.}
\label{fig:single_23}
\end{figure}

\begin{figure*}
\centering  
\includegraphics[width=1.\textwidth]{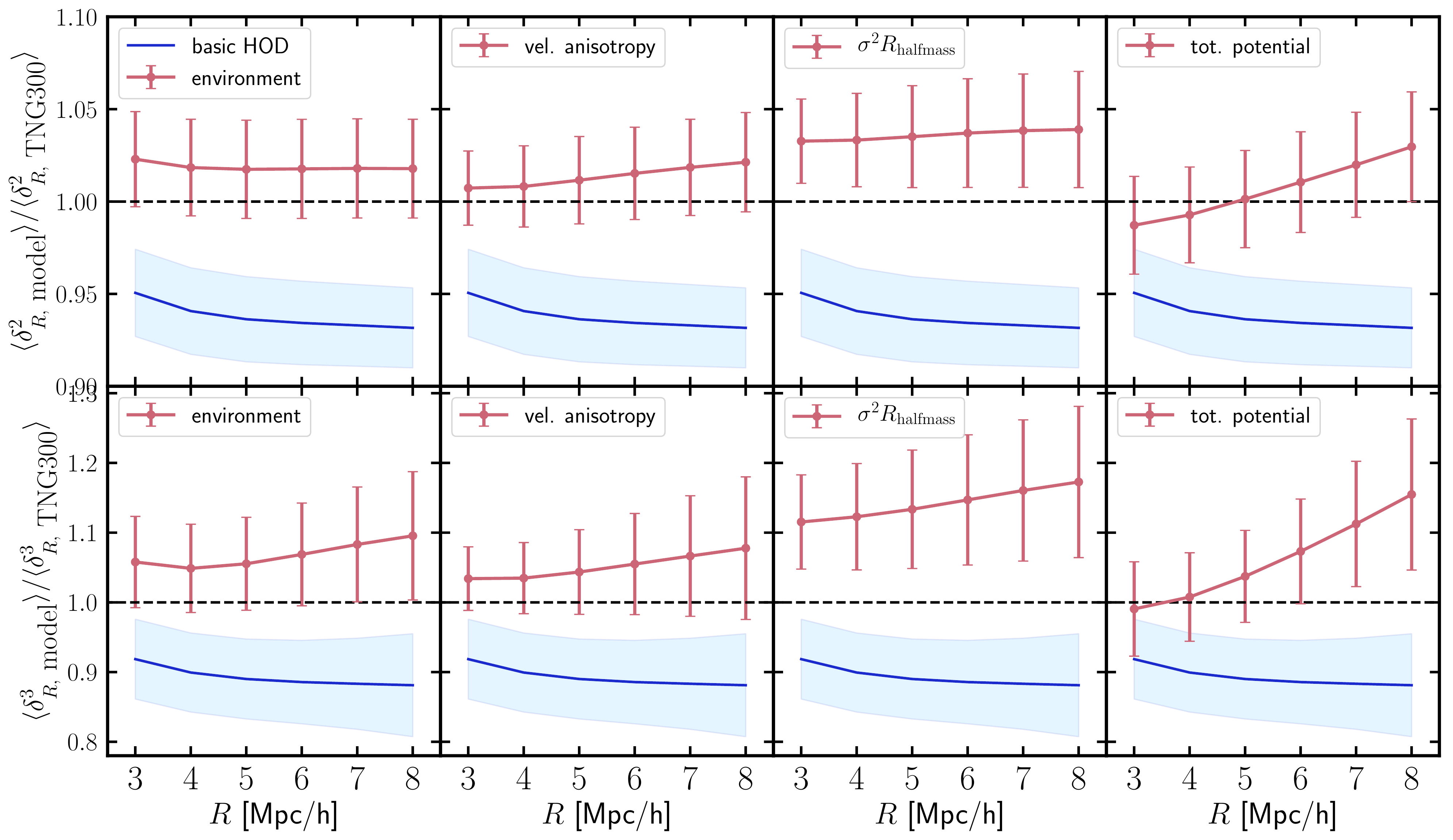}
\caption{Ratio of the second and third moment of the
smoothed galaxy density field between the 2D HOD 
galaxy sample and the ``true'' TNG300 sample
(shown in Fig. \ref{fig:single_23}). \textit{The
top panels} correspond to the ratios of the
second moments with respect to the ``true'' TNG300
population, while the \textit{bottom panels} show the ratio of their third moments. The blue shaded lines 
correspond to the ``basic'' HOD model, while
orange corresponds to the ``fitted'' 2D-HOD
models. We start smoothing at 3 Mpc$/h$,
which is around the transition scale
between the 1-halo and 2-halo terms. On
larger scales, the dominant effect comes from
large clusters, and we see that while the ``basic''
HOD model falls lower from the ``truth'', the
other 2D-HOD models seem to exhibit a better
agreement, the exception being $\sigma^2
R_{\rm halfmass}$.}
\label{fig:2nd_3rd}
\end{figure*}

The procedure we follow can be outlined as follows:
\begin{itemize}
\item Divide the TNG box into $512^3$ cubes  of side $\sim 0.4$ Mpc$/h$ and
compute the counts-in-cell density field in each as $\delta_i = N_i/{\bar N}-1$;
\item Convolve it with a Gaussian filter of smoothing 
scale $R = \{3,4,5,6,7,8\}$ Mpc/$h$ to get the smoothed density field
$\delta_R$;
\item Compute the second and third moments of the density
contrast as $\langle \delta^2_R \rangle$ and $\langle \delta^3_R \rangle$,
respectively;
\item Study how the values of the two moments change as a function of
smoothing scale for different galaxy distributions.
\end{itemize}
Fig. \ref{fig:single_23} illustrates plots of the second and third moment as a
function of the Gaussian smoothing kernel for the TNG300 (``true'') galaxies and the ``basic'' HOD model.
One can notice that the values of the moments decrease more steeply
the more smoothed the density field is, as more and 
more of the information
on intermediate scales ($\sim 1$ Mpc$/h$) gets 
erased. The discrepancy between the two models is even more pronounced in Fig. \ref{fig:2nd_3rd}, which shows their
ratio alongside the other models we test.
The error bars are
roughly constant across all smoothing scales (seen more clearly in 
Fig. \ref{fig:2nd_3rd}) as in this case, the jackknifing is
performed on the three-dimensional smoothed density field by consecutively excluding
subboxes from it (see Section \ref{sec:jack}), 
which is independent of the smoothing scale.

In Fig. \ref{fig:2nd_3rd}, we demonstrate how the second and third
moments of the smoothed galaxy density field compare with TNG300 as
a function of smoothing scale. Since the cumulants are closely
related to the two- and three-point clustering statistics, it
is not surprising (but reassuring to see)
that the ``basic'' (mass-only) HOD model falls short of capturing
the statistical behavior of the TNG300 galaxy sample. More surprising,
however, is the finding that the 2D-HOD galaxy sample with 
$\sigma^2 R_{\rm halfmass}$ as a secondary parameter also fails
that test despite being tuned to match the two-point correlation
function on large-scales (more prominently observed
in the third-moment panel). This indicates that 
conditioning on this parameter in the 2D-HOD model leads to
a difference in the distribution of galaxy clusters and
should make us cautious of using it for populating
halos.

\subsubsection{Void statistics}
Cosmic voids are large 
underdense regions with typical sizes of $10 - 100$
Mpc$/h$. They have undergone very little non-linear growth
compared with halos
and thus offer a pristine probe
for studying cosmology \citep{1978ApJ...222..784G}. 
They are sensitive to a number of effects such as
redshift space distortions,
baryon acoustic oscillations, neutrino signatures,
and the integrated Sachs-Wolde effect
\citep[e.g.]{2019MNRAS.488.4413K}. As an example,
the anisotropic galaxy distribution around
voids can be used as an Alcock-Paczynski test
\citep{1979Natur.281..358A}. 
As voids are regions of low density,
the average galaxy velocity in the vicinity of voids is directed
outwards from the void centre. This causes a distortion
of the cross-correlation of void centres and galaxy positions,
as the void interior is stretched along the line-of-sight,
while the void walls are squashed,
e.g. \cite{2019PhRvD.100b3504N}. Voids
are complementary to both galaxy clustering and
early-Universe measurements and can help
break existing degeneracies between cosmological parameters.

In this work, we analyze the void size distribution and
cross-correlation between galaxies
and voids in real space with the intention of gaining an
alternative probe of the galaxy distribution when using
different population models. The number of voids
as a function of their size can tell us whether the 
population model we employ assigns more galaxies in 
underdense regions than the hydro simulation, which
would result in a larger number of small-sized voids
at the expense of large voids. We do observe this
tendency when comparing the ``basic'' HOD model and
TNG300 but not at a significant enough level, so we do not show
the result in this paper. On the other hand, if we
overpopulate already dense regions, then the largest 
voids increase their number relative to TNG300.
Neither of these effects can be confirmed with a
sufficient level of precision, so we leave it for
future analysis with a larger hydro simulation box.
On the other hand, the location of the voids
informs us where the most galaxy-deprived regions are
located and cross-correlating those with the galaxy
position offers insight into the relationship between
large underdensities and densely clustered regions.

\begin{figure}
\centering  
\includegraphics[width=0.5\textwidth]{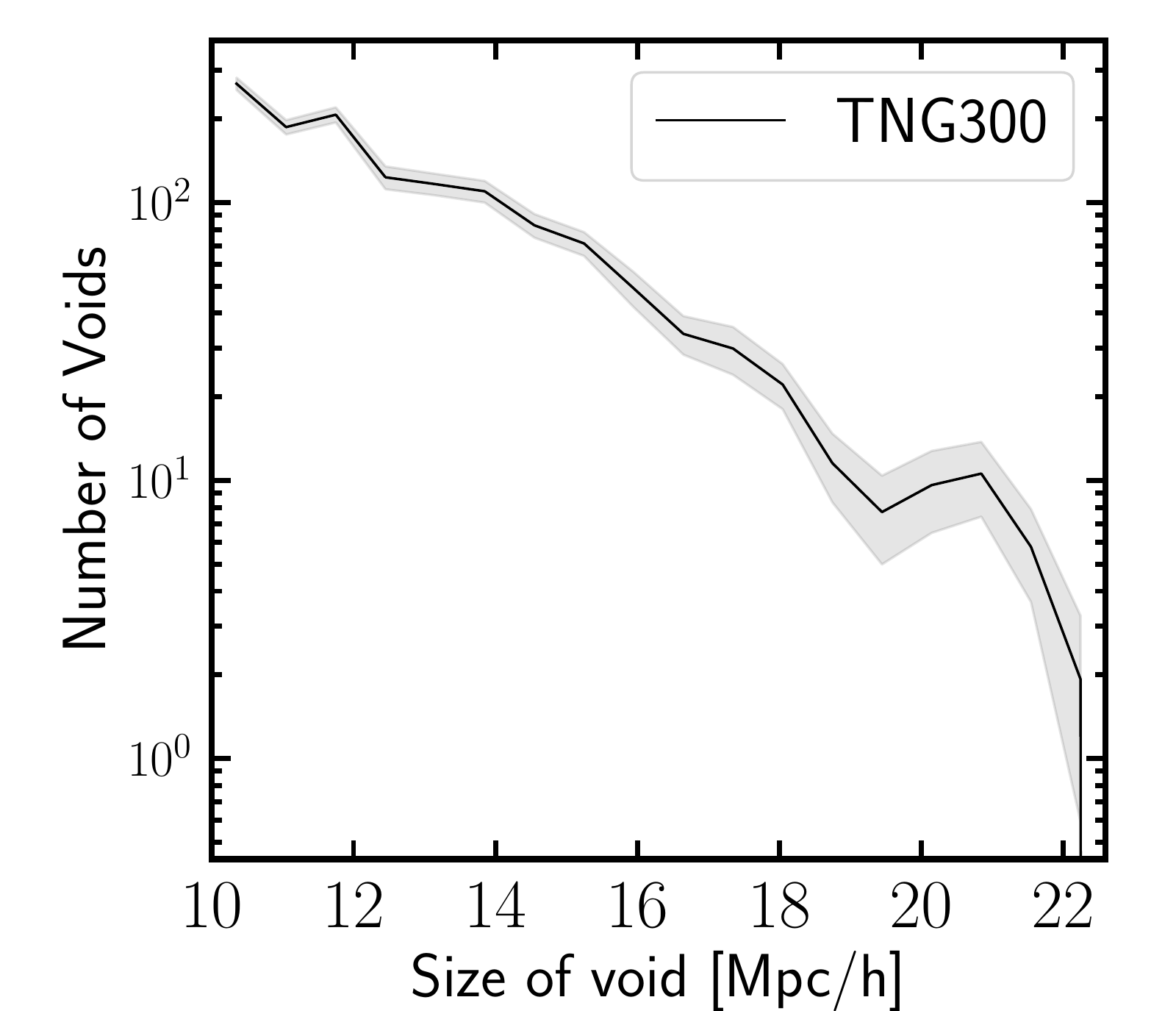}
\caption{Number of voids as a function of their size
for the ``true'' galaxies in TNG300. The error
bars are derived from jack-knifing the final
void distribution. While we do not
observe a strictly monotonic decrease (in log-space) for
the larger voids, that is most likely the case due
to their limited numbers ($\sim 10$).}
\label{fig:true_void}
\end{figure}

We have devised our own heuristic to 
infer the sizes and positions
of the largest voids in the TNG300 box.
We note that this method is not as sophisticated
as some of the already existing algorithms used for void analysis,
but since we are comparing the galaxy populations in a consistent
way, i.e. using the same void definition, the qualitative
conclusions derived are still meaningful. Our recipe is 
the following
\begin{itemize}
\item Divide the TNG box into $128^3$ cubes  of side $\sim 1.6$ Mpc$/h$;
\item Find the distance to the {\it third} nearest galaxy measured from the center of each of the cubes;
\item Order the thus obtained void candidates based on their size
in descending order and going through each object in the list,
remove all voids whose centers lie within the boundaries of
that object;
\item Record the void sizes and void centers
for each of the population scenarios of interest.
\end{itemize}
A histogram of the void sizes for the ``true'' (TNG300) galaxy sample
is shown in Fig. \ref{fig:true_void}. The smallest voids that we
find through this method are of radius of 10 Mpc$/h$, while the
largest reach about 22 Mpc$/h$. These are modest sizes for 
voids which limit the conclusions we can draw
from analyzing them. Furthermore, we work with a 
relatively small number of voids, on the order of 1000, so 
our findings are further inhibited by the void scarcity in TNG300.
The overall shape of the void size distribution agrees with
previous analyses \citep[e.g.][]{2017A&A...607A..24R}.
We do not show void size distribution comparisons
with the other population models as their deviations from TNG300 are
not statistically significant.

\begin{figure}
\centering  
\includegraphics[width=0.5\textwidth]{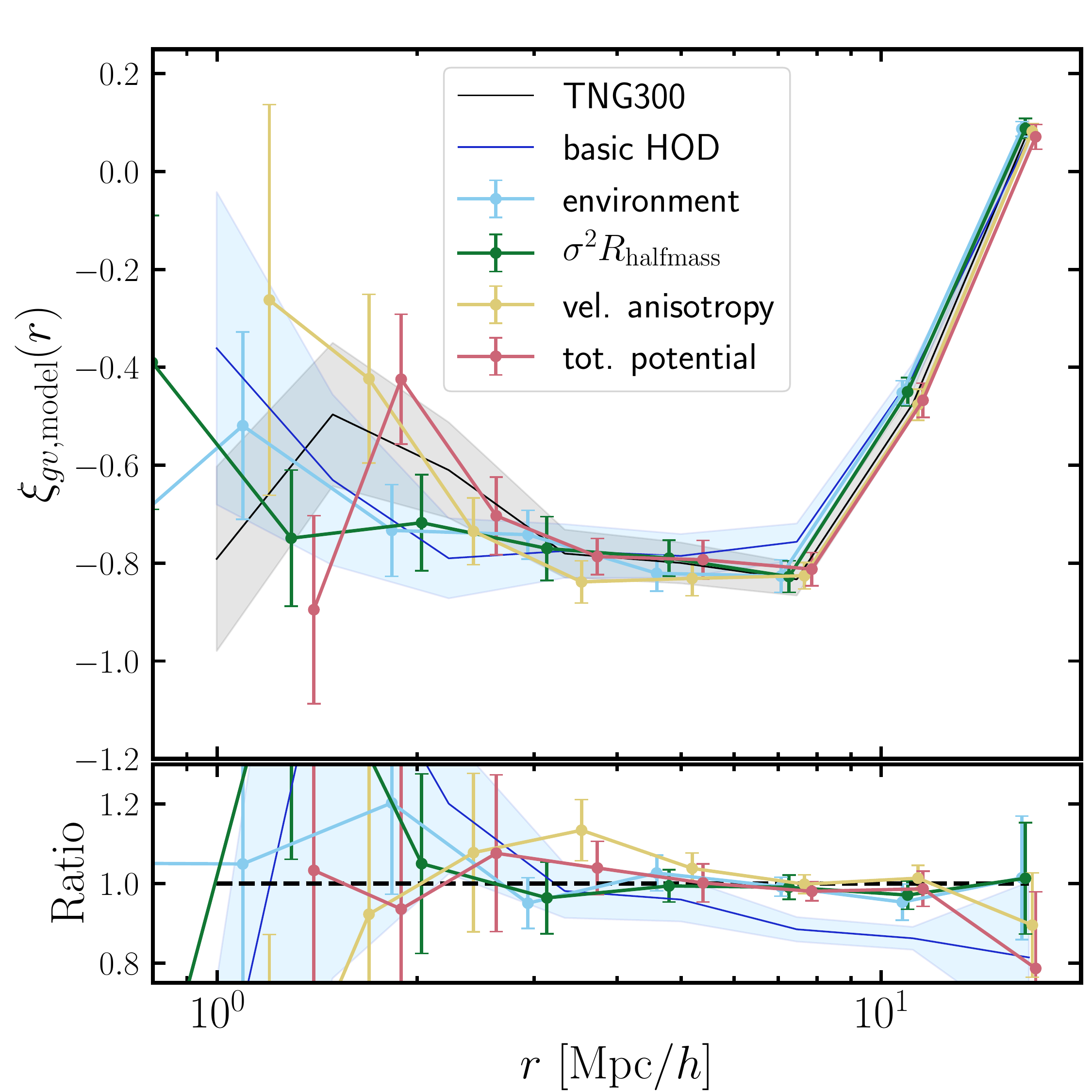}
\caption{Galaxy-void cross-correlation function
and ratio between the various HOD models
presented in this work and the ``true'' galaxy
sample. The blue shaded curves correspond to the
``basic'' (mass-only) HOD model, while the curves
with vertical error bars to the
``fitted'' 2D-HOD samples. The
first three 2D-HOD models
(i.e. $\sigma^2 R_{\rm halfmass}$) seem to be in 
good agreement with the ``true'' population,
while the basic HOD model and the total 
potential 2D-HOD distribution exhibits
more pronounced differences especially
on large scales. The voids used for this figure
are of sizes 10 Mpc$/h$ and above.} 
\label{fig:voids}
\end{figure}

Fig. \ref{fig:voids} illustrates the real-space cross-correlation
function between voids and galaxies found in the hydro simulation
(TNG300), the ``basic'' (mass-only) HOD model, and the four 2D-HOD
models considered (augmented with one of the following
secondary parameters: environment, velocity, anisotropy, 
$\sigma^2 R_{\rm halfmass}$, and total potential).
For almost all scales considered, they are negatively 
correlated. There is an upsurge around $r = 8$ Mpc$/h$,
which roughly corresponds to the smallest void size
we find. On larger scales, there is a positive 
correlation between void centers and galaxy 
positions expected to keep increasing on
even larger scales. The range over which we can study it,
however, is limited by the simulation volume
to around $r = 20$ Mpc$/h$. From the \textit{bottom panel},
which shows the ratio between the cross-correlation
function for the different models, we can see that
the mass-only HOD model performs poorly, exhibiting a
deviation of $\gtrsim 10\%$ at $r = 10$ Mpc$/h$. The
four 2D-HOD models do substantially better, on the
other hand. The largest fluctuation observed is again
when using the total potential as a secondary parameter.
This finding further cautions us against using this parameter
in population modeling.

There is also a connection between void statistics and the
moments of the density field, which stems from the fact
that the high-point moments exaggerate the underdense regions and
hence its average value, $\langle \xi^3 \rangle$, is dominated by 
the underdensities. This can be seen in practice when
comparing different population models in terms of the
resulting void size distributions and third moments. We have
done this for augmented HOD models,
where we assume perfect correlation between the secondary moment
(e.g. environment) and the galaxy occupation (i.e. $r = 1$, see Section 
\ref{sec:2dhod}), and found that they have both higher third moments
as well as more large voids.

\section{Discussion and Conclusion}

One of the main strengths of empirical approaches to modeling the galaxy-halo connection is that they
allow us to constrain the DM distribution by using observations
of the biased luminous components.
Applying these models, we can obtain mock
catalogs which are then used
to build high-precision covariance matrices for quantifying
the uncertainties in estimates of cosmological parameters. These
are particularly important for next-generation large-scale
structure experiments, such as
DESI \citep{2013arXiv1308.0847L} and {\it Euclid}
\citep{2018LRR....21....2A}. 
Empirical models considerably speed up the
construction of mock catalogs and are widely used for
forward modeling cosmological observables, 
e.g. \cite{2013MNRAS.432..743N,2015JCAP...06..015L,2016MNRAS.456.4156K}.
In addition, empirical models can complement and help improve
\textit{ab initio} models,
such as hydrodynamical and semi-analytical models of galaxy formation,
which tend to have more free parameters
to incorporate baryon physics and are considerably more expensive to run.

In this paper, we have first explored the empirical model
called subhalo abundance matching, or SHAM.
We have shown that the subhalo parameters that
exhibit the least amount of discrepancy compared with TNG300
are the subhalo peak velocity (i.e. the highest circular 
velocity it has reached throughout its history) and the
relaxation velocity (the highest circular velocity it has
reached throughout the times in which it satisfies a
relaxation criterion defined in Section \ref{sec:sham}). These parameters manage to successfully
reproduce the large-scale clustering to within 1\% and 3\%, 
respectively, for the number density considered 
($n_{\rm gal} \approx 1.3 \times 10^{-3}
\ [{\rm Mpc}/h]^{-3}$) and in particular, perform
better than the ``basic'' HOD model. However, on small scales
($\lesssim$ 1 Mpc$/h$), they exhibit significant discrepancies when 
compared with TNG300. 

To understand which subcategories of 
galaxies are represented best by the SHAM model,
we have split our sample into 5 pairs of sub-populations,
based on stellar mass, color, star-formation
rate, and hierarchical position in the halo
(see Fig. \ref{fig:sham_split}). 
We have found that the central galaxy
sub-population seems to correlate the strongest with the
large-scale distribution of the corresponding
DM-only subhalos in the SHAM catalog. Introducing scatter
into the $M_{\rm star}-V_{\rm peak}$ relationship would
decrease the clustering and thus bring the two
samples into better agreement. This result also implies 
that to achieve greater consistency, one ought to
adopt population mechanisms which treat satellites
and centrals differently. Furthermore, the high-mass, blue, and
star-forming subcategories, which are predominantly
made up of central galaxies in our sample, additionally
boost the ratio of the clustering to their SHAM catalog equivalents. 
Finally, we have seen that the late-forming galaxies in our stellar-mass-cut
selected sample are overwhelmingly classified as centrals ($\sim 95\%$),
and therefore follow a very similar trend to that observed
for the centrals.

Furthermore, we have investigated the effect of 
baryonic processes on subhalo properties,
which may contribute significantly to the bias between
the SHAM-based and  ``true'' galaxy distributions.
It is known that energetic processes such as
AGN feedback expel material from the subhalo
which causes its intrinsic properties, such as
its concentration, to vary considerably between
the dark-matter only and the hydrodynamical simulations \citep{2017MNRAS.469.1997D,2017MNRAS.472.2153P}.
We have explored how subhalos are affected across
the two simulations by matching the best resolved
several thousand galaxies in the full-physics run
with their counterparts in the DM-only run and
comparing different properties used in the SHAM
model construction such as peak mass and infall
velocity. This analysis reveals that the full-physics quantities are consistently lower than the dark matter ones
(see Fig. \ref{fig:sham_frac}),
which suggests that tidal stripping, feedback, and
galactic disk effects play a significant role
in altering these parameters, which ultimately
affects the relative ranking of the subhalos in 
full-physics and dark-matter simulations and as
such leads to a decrease in the inferred clustering
relative to the ``true'' galaxy distribution.
Even at the level of the bijective matching,
we have found that there is a significant fraction
of the objects that have no direct matches,
indicating that the inclusion of baryonic physics
changes significantly the one-halo clustering term.

Another factor which influences the extracted
properties of subhalos and halos from N-body 
simulations and thus may introduce significant
systematic issues is the choice of subhalo and
halo finding algorithms. 
Many works have recently expressed concerns about the
accuracy of the most widely used Friends-of-Friends
method, and alternatives such as the temporal
phase-space halo finder ROCKSTAR have been
viewed more favorably
\citep{2009ApJ...692..217L,2011ApJS..195....4M,2011MNRAS.415.2293K}. 
In this paper, we have not
explored the question of building population models adopting
other halo finders, but leave this for the future.
A full-merger tree analysis performed with
ROCKSTAR opens room for interesting comparisons
between the two halo finders as well as comparisons
between phenomenological models such as SHAM and
physically motivated ones such as semi-analytical
models.

Since the large-scale clustering obtained from the
IllustrisTNG 300 Mpc simulation box matches
the clustering of real galaxies
reasonably well \citep{2018MNRAS.475..676S},
the extension to the HOD model 
developed in this paper and its subsequent
statistical analysis can provide a framework for developing
improved mock catalogs in preparation for future surveys.
To mitigate the discrepancies observed in the large-scale
clustering of galaxies, we have introduced a two-dimensional
HOD (2D-HOD) model that is augmented
with a secondary halo parameter
in addition to mass. The parameters included in this extended
HOD model are the total potential energy of the halo, its
local environment, the velocity anisotropy of the particles
within it, and the virialized mass
measured as $\sigma^2 R_{\rm halfmass}$.
There is one free parameter in this model, $r$, measuring the
amount of correlation between halo occupancy and the secondary
parameter within each 5\% mass bin. We choose its value so as
to reconcile the galaxy auto-correlation function on large
scales ($1 - 20$ Mpc$/h$) to within $\sim 1$\%,
the required precision of current cosmological efforts. 
We manage to do this for all 
4 parameter choices listed above, 
the only exception being the total potential 
energy, where the galaxy bias is scale-dependent and does not
approach constant behavior on 
large scales as in the other cases
(see Fig. \ref{fig:bias_corr}).

We have explored different statistical properties of
the 2D-HOD galaxy catalogs and shown that despite
being fitted to match only the two-point clustering
of galaxies, two of our parameter choices
(environment and velocity anisotropy) demonstrate
an excellent agreement with TNG300 for all the examined statistical
observables. We believe this to be a significant finding
that merits further exploration.

We have shown that the
stacked lensing of all the 2D-HOD samples is
in very good agreement with TNG300, the worst choice for a 
secondary parameter being again the total 
potential energy of the halo, which might have been anticipated
by our results concerning the auto-correlation
function. We argue that the
discrepancy observed in cross-correlation measures in
real space ought to be half 
that found in their auto-correlation.
Furthermore, our results indicate that the linear bias
approximation works well 
on scales larger than 10 Mpc$/h$,
where the correlation coefficient approaches 1
and the galaxy bias is roughly constant. This suggests 
that on these scales, baryon physics does not affect the
galaxy distribution, and the dominant source that governs
the galaxy distribution is gravity. We have next examined
the question of whether our 2D-HOD galaxy catalogs which were 
constructed so as to match the 
two-point correlation function provide an equally
good match to cross-correlation measures between the matter
and galaxy distributions such as galaxy-galaxy lensing (Fig. 
\ref{fig:gal_lens}) and the correlation coefficient 
(see Fig. \ref{fig:bias_corr}). 

The other statistical tools we have employed in this 
work are cosmic void statistics and cumulants of the smoothed
galaxy overdensity, both of which contain higher-point
information than the galaxy auto-correlation function. 
The analysis of the void size distribution does not yield
a statistically significant difference between the different
models, which is most likely due to our small-number
and small-radius limitations ($\sim$ 1000 voids of 
maximum radius of 22 Mpc$/h$). Studying the void-galaxy cross-
correlation function provides us with more illuminating
insights that allow us to discriminate better between the
different population models. In particular, the 2D-HOD
model augmented with the total potential secondary parameter
exhibits a larger discrepancy compared with the rest of the
parameters. A caveat of our void definition
is that we consider only spherically shaped voids.
With a larger hydro simulation box, we could potentially
begin to see more prominent differences in the different
population mechanisms. 

For our analysis of the cumulants of the
density field, we have shown the values of the second and
third moments as a function of the smoothing 
scale ($3 - 8$ Mpc$/h$). Of particular
interest is the scale of clusters, $\sim 3$ Mpc$/h$, for which
we have demonstrated that the 2D-HOD 
samples are roughly consistent with
the ``true'' TNG sample, while the ``basic''
(mass-only) HOD model seems
to deviate by more than $1\sigma$ (see Fig. \ref{fig:2nd_3rd}). The worst performing
secondary parameter choice is the virial mass 
measure $\sigma^2 R_{\rm halfmass}$. Of the 4
secondary parameters considered, across all statistics
measures tried out in this paper, the best performing
ones are the local environment (an extrinsic
halo property) and the velocity
anisotropy (an intrinsic halo property).
We consider this a non-trivial result, since the 2D-HOD models
have only been fitted to match the two-point clustering,
and yet, two of resulting samples exhibit a remarkable consistency
with TNG300 for all statistical probes considered in this paper.

In the near future, even larger hydrodynamical
galaxy formation simulations will be available, which
will be extremely beneficial for expanding our knowledge
of the relationship between galaxies and their dark 
matter halos. Once such data sets become available,
we plan to test and validate the results obtained with TNG300
as well as improve the empirical population models used
for creating mock catalogs. Thanks to the substantially 
larger number of galaxies contained in these larger volume runs,
such simulations will enable us to capture the large-scale
behavior even better by possibly introducing a multidimensional
approach to the HOD model. Such an endeavor could potentially
open the door for creating efficient galaxy
population models that recover
the galaxy clustering on large scales with subpercent
precision and thus bridge important
gaps in light of future galaxy surveys. 

\section*{Acknowledgements}
We would like to thank Christina Kreisch, Masahiro Takada,
Charlie Conroy and David Spergel for their valuable insights and
enlightening discussions.
DJE is supported by U.S. Department of Energy 
grant DE-SC0013718 and as a Simons Foundation Investigator.

%%%%%%%%%%%%%%%%%%%%%%%%%%%%%%%%%%%%%%%%%%%%%%%%%%



\bibliographystyle{mnras}
\bibliography{refs} % if your bibtex file is called example.bib

%%%%%%%%%%%%%%%%%%%%%%%%%%%%%%%%%%%%%%%%%%%%%%%%%%

%%%%%%%%%%%%%%%%% APPENDICES %%%%%%%%%%%%%%%%%%%%%

%\appendix

%%%%%%%%%%%%%%%%%%%%%%%%%%%%%%%%%%%%%%%%%%%%%%%%%%

% Don't change these lines
\bsp	% typesetting comment
\label{lastpage}
\end{document}